\documentclass{aa}

\usepackage{graphicx}
\usepackage{natbib}
\usepackage{scalerel}

\usepackage[table]{xcolor}

\usepackage{txfonts}
\usepackage{subfig}
\usepackage[ruled,linesnumbered, onelanguage]{algorithm2e}
\usepackage{float}
\floatstyle{ruled}
\newfloat{algorithme}{hbt}{alg}
\floatname{algorithme}{Algorithme}
\floatstyle{boxed}
\usepackage{graphicx}
\usepackage{tikz}

\definecolor{lightcyan}{rgb}{0.97, 0.91, 0.81}

\usepackage[pdfencoding=auto,psdextra]{hyperref}
\hypersetup{
    colorlinks=true,
    linkcolor=blue,
    filecolor=magenta,      
    urlcolor=blue,
    citecolor=blue
}
\makeatletter
\renewcommand*\aa@pageof{, page \thepage{} of \pageref*{LastPage}}
\makeatother

\usepackage[utf8]{inputenc}

\usepackage[switch, modulo]{lineno}

\usepackage{euclid}

\begin{document}

\title{New methods to improve the decontamination of slitless spectra}

\author{M.~Bella\inst{1}\thanks{Corresponding author, \email{mbella@irap.omp.eu}}
	\and S.~Hosseini\inst{1}
	\and T.~Contini\inst{1}
	\and H.~Saylani\inst{2}}

\institute{
	IRAP, Université de Toulouse, CNRS, CNES, 14 Av. Edouard Belin, 31400 Toulouse, France
	\and
	MatSim, Faculté des Sciences, Université Ibnou Zohr,	BP 8106 Cité Dakhla, Agadir, Maroc
}

\date{}

   \abstract{
This paper proposes four new methods to decontaminate spectra of stars and galaxies
resulting from slitless spectroscopy used in many space missions such as  \Euclid. These methods are based on two distinct approaches and simultaneously take into account multiple  dispersion directions of light.
The first approach, called the local instantaneous approach, is based on an approximate linear instantaneous model.
The second approach, called the local convolutive approach, is based on a more realistic convolutive model that allows simultaneous decontamination and deconvolution of spectra.
For each approach, a mixing model was developed that links the observed data to the source spectra. This was done either in the spatial domain for the local instantaneous approach or in the Fourier domain for the local convolutive approach. Four methods were then developed to decontaminate these spectra from the mixtures, exploiting the direct images provided by  photometers. Test results obtained using realistic, noisy, \Euclid-like data confirmed the effectiveness of the proposed methods.}


   \titlerunning{New methods to improve the decontamination of spectra}
   \authorrunning{M. Bella et al.}
   
\maketitle
   
\section{Introduction}
\label{introduction}

The aim of source separation is to recover a set of unknown signals, called sources, from a set of mixtures of these sources, called observations. When this estimation is done with minimal a priori information about the sources and the mixing parameters, it is referred to as Blind Source Separation (BSS) \citep{Comon,Hyvarinen,Deville16}. BSS is receiving increasing attention due to the diversity of its fields of application, including audio processing \citep{audio1,BELLA2}, Earth observation \citep{hyp1,hyp2}, biomedical applications \citep{bio1,bio2} and astronomy \citep{astro2,astro1}. 
In this paper, we address a new application of source separation concerning the decontamination of spectra resulting from slitless spectroscopy, a technique commonly used in various space missions such as the James Webb Space Telescope (JWST) \citep{JWST}, the
  Hubble Space Telescope (HST)  or in the  \Euclid space mission \citep{EuclidSkyOverview}. 
Our methods are specifically designed for \Euclid-like configurations, but can also be adapted to other telescopes that use slitless spectroscopy, like JWST and HST.

\Euclid is a space telescope of the European Space Agency successfully launched on July 1, 2023 \citep{EuclidSkyOverview}.  Its primary objective is to understand the nature of dark energy and how it is responsible for the accelerating expansion of the Universe.
\Euclid will create a three-dimensional map of the Universe, with time as the third dimension. This map will represent
millions of galaxies, revealing the evolution of the universe over time and providing crucial information about the nature of its accelerating expansion  \citep{Laureijs11,EuclidSkyOverview}.
To achieve this goal, \Euclid is equipped with a slitless near-infrared spectro-photometer  called NISP (Near Infrared Spectrometer Photometer) \citep{Costille,Schirmer2022,EuclidSkyNISP}.
NISP will measure the spectra of millions of galaxies. These spectra will then be analyzed to estimate the spectroscopic redshifts of the observed galaxies, focusing in particular on the $H_\alpha$ line, which is usually the strongest emission-line in star-forming galaxies and thus the easiest to detect in the spectrum of a galaxy.
The NISP instrument provides both direct non-dispersed images from a photometer and dispersed images (corresponding to spectra) from a slitless spectrograph for all objects within the field of view. Photometry in NISP is performed using three filters, resulting in three near-infrared photometric images in the three bands  $Y$, $J$, and $H$ \citep{EuclidSkyNISP}.

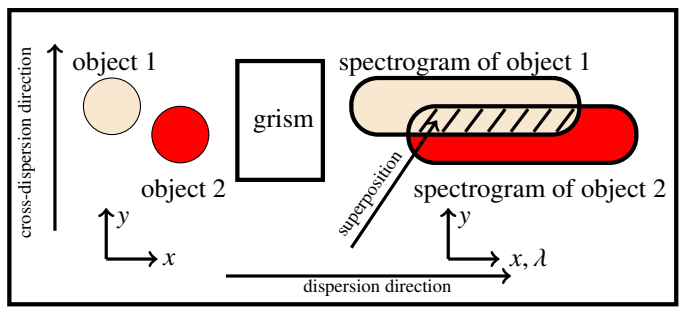
\begin{figure}[h!]
	\begin{center}
		\begin{tikzpicture}[scale=0.75]
		\draw [black,fill=lightcyan] (1.8,1.5) circle [radius=0.5];
		\draw [black,fill=red] (3.0,1) circle [radius=0.5];
		\node at (3.0,0) {{ object 2}};
		\node at (1.8,2.25) {{ object 1}};
		\node at (2.0,-0.5) {$y$};
		\node at (2.8,-1.2) {$x$};
		\node at (7.99,-0.5) {$y$};
		\node at (9.1,-1.2) {$x, \lambda$};			
		\node at (9.25,0) {{ spectrogram of object 2}};
		\node at (8.0,2.25) {{spectrogram of object 1}};
		\draw [black, ultra thick] (4.0,0.2) rectangle (5.5,2.3);
		\draw [lightgray,fill=red,rounded corners=10] (7.0,0.5) rectangle (11.0,1.5);
		\draw [green,fill=lightcyan,rounded corners=10] (6,1) rectangle (10,2);
		\draw [black,ultra thick,rounded corners=10](7.0,0.5) rectangle (11.0,1.5);
		\draw [black,ultra thick,rounded corners=10] (6,1) rectangle (10,2);
		\node at (4.75,1.25) {{ grism}};
		\draw [black,very thick,->]  (3.8,-1.5) -- (8.8,-1.5);
		\draw [black,very thick,->]  (0.8,-0.7) -- (0.8,2.6);
		\draw [black,very thick,->]  (1.7,-1.2) -- (1.7,-0.3);
		\draw [black,very thick,->]  (1.7,-1.2) -- (2.6,-1.2);
		\draw [black,very thick,->]  (7.7,-1.2) -- (7.7,-0.3);
		\draw [black,very thick,->]  (7.7,-1.2) -- (8.6,-1.2);				
		\draw [black,very thick,->]  (6.0,-1.0) -- (7.5,1.25);
		\draw [black,very thick,-]  (7.2,1.05) -- (7.5,1.45);
		\draw [black,very thick,-]  (7.6,1.05) -- (7.9,1.45);
		\draw [black,very thick,-]  (8.0,1.05) -- (8.3,1.45);
		\draw [black,very thick,-]  (8.4,1.05) -- (8.7,1.45);
		\draw [black,very thick,-]  (8.8,1.05) -- (9.1,1.45);
		\draw [black,very thick,-]  (9.2,1.05) -- (9.5,1.45);	
		\draw [black,very thick,-]  (9.6,1.05) -- (9.9,1.45);															
		\node [rotate=90] at (0.3,0.8) {{ {\scriptsize cross-dispersion direction}}};
		\node [rotate=55] at (6.3,-0.1) {{ {\scriptsize superposition}}};
		\node at (6.4,-1.7) {{ {\scriptsize dispersion direction}}};
		\draw [black, ultra thick] (-0.0,-2.0) rectangle (11.7,3.2);
		\end{tikzpicture}
	\end{center}
	\caption{Contamination of the spectra of neighboring objects at the output of the grism.}
	\label{contamination2}
\end{figure}

Spectroscopy is performed using grisms, which are a combination of prisms and diffraction gratings \citep{Grism}. Each grism disperses the different wavelengths of the received light.  In practice, several grisms can be used so that each grism disperses the received light in a specific dispersion direction \citep{EuclidSkyNISP}. The output of the grism is a multi-row image, hereafter referred to as a {\it spectrogram}.
Spectroscopy is usually performed with a slit that acts as a filter, allowing light to pass only from a small region of the sky containing a very limited number of objects. As a result, spectra from different sources can only be superposed if two or more objects are within the window defined by the slit. 
In contrast, slitless spectroscopy does not use a slit, which means that there is no filtering of sources, which allows a large number of objects to be observed in a single acquisition.
Fig. \ref{contamination2} illustrates an example of two direct images of nearby astronomical objects, accompanied by their respective spectrograms after being dispersed by the grism. This figure clearly shows that  slitless spectroscopy used  leads to the superposition of spectrograms of astronomical objects (galaxies and stars). 
This superposition phenomenon is called contamination.
It is important to note that the spectrogram of an object of interest can often be affected by several neighboring objects, which can lead to incorrect redshift estimates \citep{Laureijs11}. Therefore, it is essential to use an effective decontamination method to minimize errors in redshift measurements \citep{HOSSEINI2020}.

\begin{figure*}[h!]
	\begin{center}
		\begin{tikzpicture}[scale=0.9]
		\draw [red,rotate=+4,fill=lightgray, ultra thick,rounded corners=10] (-4.0,5.2)  rectangle (1.7,5.7);
		\draw [red,rotate=-4,fill=lightgray, ultra thick,rounded corners=10] (1.1,5.4) rectangle (6.8,5.9);
		\draw [red,rotate=0,fill=lightgray, ultra thick,rounded corners=10] (-4.4,5.8) rectangle (1.3,6.3);
		\draw [red,rotate=0,fill=lightgray, ultra thick,rounded corners=10] (1.5,5.8) rectangle (7.2,6.3);
		
		\node at (-1.5,6.5) {{ Direction 180$^{\circ}$}};
		\node at (4,6.5) {{ Direction 0$^{\circ}$ }};
		\node at (-1.5,4.75) {{ Direction 184$^{\circ}$ }};
		\node at (4,4.75) {{ Direction -4$^{\circ}$ }};
		\draw[-latex] (-4.4,6.0) arc (120:240:0.5) ;
		\draw[-latex] (7.2,6.0) arc (120:-120:0.5) ;		
		\node at (-5.95,5.5) {{Rotation 4$^{\circ}$}};		
		\node at (9.3,5.5) {{ Rotation -4$^{\circ}$}};										
		\end{tikzpicture}
	\end{center}
	\caption{\Euclid's observation strategy.}
	\label{strategyK}	
\end{figure*}
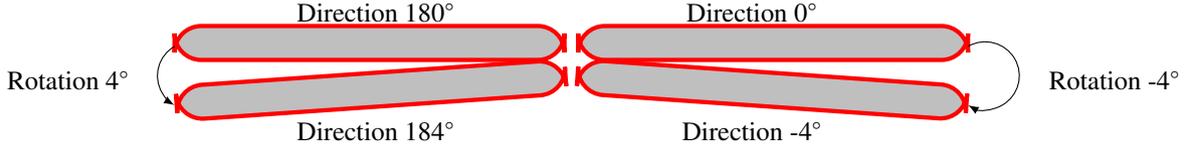

To improve the efficiency of decontamination methods,
multiple spectrograms are usually generated in different dispersion directions for the same field of view.
In fact, when moving from one dispersion direction to another, the spectrum of an object of interest is often contaminated by different spectra. As a result, the simultaneous use of multiple dispersion directions provides a better estimate of the spectrum of the object of interest \citep{Scaramella-EP1}. For example, the
\Euclid's observation strategy consists of generating spectrograms in four different dispersion directions, namely $0$, $180$, $184$ and $-4$ degrees. To achieve this, as illustrated in Fig \ref{strategyK}, two grisms with opposite dispersion directions ($0$ and $180$ degrees) are used, each of which can be rotated by 4 degrees \citep{Scaramella-EP1}.

Spectra decontamination in slitless spectroscopy can be seen as a source separation problem.
In this application, source separation can be achieved by either a local or a global approach. 
The main difference between these two approaches is that the local approach decontaminates objects one by one, while the global approach decontaminates all objects in a predefined zone simultaneously. Moreover, BSS methods have been employed in the literature to address various types of mixing models. In this paper, we focus on two distinct types: the linear instantaneous model and the convolutive model. For the linear instantaneous model, the observed signal at each wavelength is represented as a linear combination of the sources  at the same detector position. In contrast, in the convolutive model, the observed signal at each wavelength  is the result of the convolution of the sources with wavelength-dependent filters.

Since the treated spectra are neither mutually independent nor sparse, classical BSS methods based on independent component analysis \citep{Comon,Hyvarinen,Comon2} and sparse
component analysis \citep{Survey2005,BELLA_access,Gribonval200} cannot be used.
For this reason, most existing BSS methods for decontaminating slitless spectroscopy spectra exploit the positivity of the data \citep{Selloum2015,Selloum2016,HOSSEINI2020}.
However, these methods have several limitations:

\begin{enumerate}
	\item They are mainly based on Non-negative Matrix Factorization (NMF) \citep{Lee1999}, which is known to be very sensitive to
	initialization, and do not guarantee the uniqueness of the solution.
	\item They generally suffer from slow convergence.
	\item In practice, they offer modest performance. 
	\item They do not use all the information available, in particular the existence of direct images provided by the photometers. 
\end{enumerate}
On the other hand, a basic slitless decontamination method for the HST was proposed in
 \citep{Basic}. This method uses only one dispersion direction at a time and is based on estimating the continuum of the contaminants in the spectrogram of the object of interest by exploiting direct images of the contaminants. The estimated contaminant contributions are then summed to create a 2D contaminant image, which is subsequently subtracted from the spectrogram of the object of interest to obtain its uncontaminated spectrum. However, this method can only decontaminate the continuum component of the spectrogram and  cannot remove emission lines from contaminating objects nor unmodeled contaminants such as hot pixels  \citep{Basic}.
	Unlike this basic decontamination method, the {\it GRIZLI} pipeline \citep{brammer2019} employs a more robust approach. While it similarly creates and subtracts contaminant models from direct images, {\it GRIZLI} refines the process by iteratively optimizing the contaminant's spectral shape. This allows for the precise subtraction of both continuum and emission-line contributions from neighboring objects \citep{momcheva20163d}. Nevertheless, the {\it GRIZLI} approach has a fundamental limitation. It cannot model contaminants that are undetected in the reference direct images. These contaminants remain a source of residual contamination.		
A distinct, matrix-based approach, known as {\it LINEAR}, was proposed by Ryan et al \citep{Linear}. Unlike the model-subtraction methods described above, {\it LINEAR} operates on a fundamentally different principle. Rather than constructing and subtracting individual source models, it frames the entire observed spectrogram as a large, sparse system of linear equations. In this system, each pixel is represented as a linear combination of contributions from every source in the field. Solving this system via regularized least squares enables the simultaneous extraction of all individual spectra. However, this method is highly computationally intensive, and its solution can be very sensitive to noise and the presence of unmodeled sources.\\

Due to the above-mentioned limitations of existing separation methods in slitless spectroscopy, this paper proposes four new separation methods adapted to \Euclid-like spectra. These methods can also be used in other slitless spectroscopy applications and are based on two different local approaches. The first approach, called the local instantaneous approach, is based on a linear instantaneous model.  The second, called the local convolutive approach, is based on a more realistic convolutive model that allows simultaneous decontamination and deconvolution of spectra.
The proposed methods provide several practical advantages compared with model-subtraction techniques (basic, {\it GRIZLI}) and global inversion frameworks such as {\it LINEAR}. In particular:

\begin{enumerate}
	\item Parallelism: The local nature of our models allows them to be run in parallel across many processors. This means we can decontaminate a large number of objects at the same time, unlike global approaches such as {\it LINEAR}, which require solving a single large system for the entire field.
	
	\item Robustness to unmodeled sources: Our methods can mitigate contamination from faint or undetected sources in the reference images, addressing a fundamental and persistent limitation of other techniques.
	
	\item Multi-directional exploitation: Our models explicitly and simultaneously use the four available dispersion directions. This integrated approach improves decontamination accuracy compared to methods that process each direction independently.
\end{enumerate}

The proposed methods are partly described in  our two conference papers \citep{BELLA3,BELLA_ICASSP}.  The present paper provides a more complete description of these two approaches, as well as additional experimental results.

The remainder of this article is organized as follows.
 Sect. \ref{models} presents  the data models used in our work, as well as the   two resulting mixing models.
  Sect. \ref{proposed} presents in detail the  proposed decontamination methods.
 Sect. \ref{test}  evaluates the performance of these methods. Finally, Sect. \ref{Conclusion} provides a conclusion and perspectives for our work.

\section{Data models}
\label{models}

In this section, we present the physical model used in our work, and two resulting mixing models that link the observed data to the source spectra. 
The first model is an approximate linear instantaneous model which will be used  in our  instantaneous methods presented in Sec. \ref{sec:LI method}. The second model is a   convolutive model  which will be used  in our  convolutive method described in Sec. \ref{conv_meth}.

\subsection{Observation model}
\label{model obs}
Assuming that the spectral distribution of any astronomical object (galaxy or star) with index $j$ is uniform at every point on the object, the light intensity of that object at a given point with coordinates $(x,y)$ and wavelength $\lambda$ can be modeled as follows
\begin{equation}
q_j(x,y,\lambda)=s_j(\lambda) \, f_j(x , y),
\label{model1}
\end{equation}
where $s_j(\lambda)$ is the spectrum of the object and $f_j(x,y)$ is its spatial intensity profile.

The modeling of the light intensity in Eq. (\ref{model1}) must take into account the convolution of the observed object with the Point Spread Function (PSF) of the instrument. If we assume that the PSF is independent of the wavelength\footnote{In practice, the PSF varies slightly with wavelength. However, this variation is not considered in this work.}, the new intensity of the observed object can be defined by the following equation
\begin{equation}
\begin{aligned}[b]
w_j( x , y , \lambda)&=q_j(x,y,\lambda)* h( x , y )\\
&=s_j (\lambda ) [f_j ( x, y )* h( x , y )],
\end{aligned}
\label{model2}
\end{equation}
where $h$ represents the PSF of the instrument, and ``$*$'' is the convolution operator.

If we assume that the light emitted by the object described in Eq. (\ref{model2}) is then dispersed by a grism in the horizontal direction $x$, we obtain a 2D dispersed image at the grism output for an object with index $j$, which corresponds to the spectrogram of this object \citep{HOSSEINI2020,Freudling}:
\begin{equation}
t_j( x , y )=\int_{\Omega_\lambda}^{}w_j( x-D(\lambda) , y , \lambda)\;\mathrm{d}\lambda,
\label{modeli3}
\end{equation}
where $\Omega_\lambda$ is the wavelength range covered by the grism, and $D(\lambda)$ is the grism dispersion function, which is the shift on the detector relative to a reference position as a function of wavelength in the dispersion direction $x$ \citep{HOSSEINI2020}.

Assuming that the grism dispersion function is expressed in the  linear form $D(\lambda)=\alpha \lambda+\beta$ (where $\alpha$ and $\beta$ are two known coefficients), Eq. (\ref{modeli3}) becomes
\begin{equation}
\begin{aligned}[b]
t_j( x , y )&=\int_{\Omega_\lambda}^{}w_j( x-\alpha \lambda-\beta , y , \lambda)\;\mathrm{d}\lambda\\
&=\int_{\Omega_\lambda}^{}I_j( x-\alpha \lambda-\beta , y)\;s_j(\lambda)\;\mathrm{d}\lambda,
\end{aligned}
\label{modeli}
\end{equation}
where $I_j(x,y)=f_j( x,y)* h( x,y)$ is the object profile with index $j$ convolved with the PSF of the instrument.

We note that the model (\ref{modeli}) is a modified form of a convolution and represents the most accurate model among those used in our work. In Sect. \ref{sec_MLI} we present a simplified model of (\ref{modeli}).

\subsection{Local instantaneous mixing model}
\label{sec_MLI}
In the local instantaneous approach, the model in Eq. (\ref{modeli}) is simplified by assuming the following separability assumption
\begin{equation}
I_j(x, y) = I_{j1}(x)I_{j2}(y).
\label{sepa}
\end{equation}
This assumption means that the function $I_j(x,y)$ is separable into $x$ and $y$.
According to \citep{Selloum2016,HOSSEINI2020}, this assumption is more realistic for small or point-like objects, as well as for circular and elliptical objects oriented in the $x$ or $y$ directions. However, for other objects, an approximation error that depends on the shape of the object is introduced in our model \citep{Selloum2016,HOSSEINI2020}.
Taking into account the assumption (\ref{sepa}), Eq. (\ref{modeli}) becomes in the dispersion direction $d_i$\footnote{
	It should be noted that, even if Eq. (\ref{modeli}) is obtained for dispersion in the horizontal direction $x$, it remains valid for all dispersion directions $d_i\in{ 0^\circ, 180^\circ, 184^\circ, -4^\circ}$ after rotating the spectrograms for each direction according to their respective angles.} 
\begin{equation}
t^{{(d_i)}}_j(x, y) =I_{j2}^{{(d_i)}}(y)\int_{\Omega_\lambda}^{}I_{j1}^{{(d_i)}}( x-\alpha \lambda-\beta)\;s_j(\lambda)\;\mathrm{d}\lambda.
\label{sep2}
\end{equation}

The observed value for a pixel with index ${\bf p}=[n, m]$ in the dispersion direction $d_i$ is expressed as follows

\begin{equation}
o^{{(d_i)}}_j({\bf p})=o^{{(d_i)}}_j(n, m) = \int\int_{(x,y) \in \Omega_{\bf p}} t^{{(d_i)}}_j(x, y) \,dx\,dy, 
\label{forme_mm}
\end{equation}
where $\Omega_{\bf p}$ is the area of pixel ${\bf p}$.
If we replace $t^{{(d_i)}}_j(x, y)$ by its definition (\ref{sep2}), we get
\begin{equation}
o_j^{{(d_i)}}({\bf p})=o_j^{{(d_i)}}( n , m )=a^{{(d_i)}}_j(  m )e^{{(d_i)}}_j( n),
\label{mod loc}
\end{equation}
with:
\begin{equation}
a_j^{{(d_i)}}( m )=\int_{y|(x,y) \in \Omega_{\bf p}} I_{j2}^{{(d_i)}}(y) \, dy,
\label{mod loc3}
\end{equation}
and 
\begin{equation}
e_j^{{(d_i)}}( n )=\int_{x|(x,y) \in \Omega_{\bf p}}\int_{\lambda\in\Omega_\lambda} I_{j1}^{{(d_i)}}( x-\alpha \lambda-\beta)\;s_j(\lambda)\;\mathrm{d}\lambda \, dx.
\label{mod loc4}
\end{equation}

Note that the value of $e_j^{{(d_i)}}(n)$ is independent of the vertical coordinate of the pixels $m$, and that the value of $a_j^{{(d_i)}}(m)$ is independent of the horizontal coordinate of the pixels $n$.

Now consider a rectangular area containing the observed spectrogram of an object of interest, in the dispersion direction $d_i$. 
Each pixel of this spectrogram receives light from $N_i$ objects (the object of interest and $N_i-1$ contaminants). In this case, the observed value in each pixel of this spectrogram can be expressed as follows

\begin{equation}
o^{{(d_i)}}({\bf p})=\sum_{j=1}^{N_i}o_j^{{(d_i)}}( n , m )=\sum_{j=1}^{N_i}a^{{(d_i)}}_j(  m )e^{{(d_i)}}_j( n).
\label{mod loc_ins}
\end{equation}

The mixing model (\ref{mod loc_ins}) can then be expressed in the following linear instantaneous form

\begin{equation}
{\bf X}^{{(d_i)}}={\bf A}^{(i)} {\bf E}^{(i)},
\label{MLI_f}
\end{equation}
where ${\bf X}^{{(d_i)}}$ is the observation matrix of size $M_i \times K$ whose element $(m, n)$ is $o^{{(d_i)}}(n,m)$,  $M_i$ represents the number of rows in the  cross-dispersion direction associated with the spectrogram of the object of interest and $K$ is the number of spectral bands, which is considered identical for all dispersion directions. 
${\bf A}^{(i)}=\left[ {\bf a}_s^{(i)}|{\bf A}_c^{(i)}\right] $ is the mixing matrix of size $M_i \times N_i$, defined by

\begin{equation}\label{mixing_matrix}
\begin{array}{r@{\,}l}
& 
\begin{matrix}
\mspace{7mu}\overbrace{\rule{1.5cm}{0pt}}^{{\bf a}_s^{(i)}} &     
\overbrace{\rule{3.25cm}{0pt}}^{{\bf A}_c^{(i)}}
\end{matrix}
\\
{\bf A}^{(i)} = & 
\left( \begin{array}{c|ccc}
a_{1}^{(d_i)}(1)  & a_{2}^{(d_i)}(1) & \ldots & a_{{N_i}}^{(d_i)}(1) \\
\vdots & \vdots    & \ddots  & \vdots \\
a_{1}^{(d_i)}(M_i)  & a_{2}^{(d_i)}(M_i) & \cdots& a_{{N_i}}^{(d_i)}(M_i)
\end{array}\right),
\end{array}
\end{equation}
where ${\bf a}_s^{(i)}=[a_{1}^{(d_i)}(1), ..., a_{1}^{(d_i)}(M_i)]^T$ is the mixing vector of the object of interest, $T$ denotes the transpose, $N_i-1$ is the number of contaminants considered in the direction $d_i$, and  
${\bf A}_c^{(i)}$ is the contaminant mixing matrix.
 As for ${\bf E}^{(i)}$, it is the matrix of sources of size $N_i \times K$, where the first row contains the vector ${\bf e}_1$, corresponding to a convolved version of the spectrum of the object of interest, while the other rows contain convolved versions of the spectra of contaminants in the direction $d_i$.

As mentioned in Sect. \ref{introduction}, we have four observations for each object of interest, corresponding to the four dispersion directions ($0$, $180$, $184$, and $-4$ degrees). 
Assuming that the spectrum of the object of interest remains the same in all dispersion directions, we can combine these four observations to improve the estimation of its spectrum.
However, before combining the spectrograms from these four directions, some pre-processing steps are necessary. First, it is necessary to rotate ${\bf X}^{(180)}$, ${\bf X}^{(184)}$, and ${\bf X}^{(-4)}$ by $180$, $184$, and $-4$ degrees, respectively, so that they are aligned with the horizontal direction \citep{BELLA3,BELLA_ICASSP}.  This preprocessing is essential to ensure that all wavelengths from all dispersion directions of the object of interest are horizontally aligned to facilitate their simultaneous exploitation.

Secondly, it is essential to re-sample (without changing the sampling rate) the four spectrograms in the dispersion direction to further align them in terms of wavelength. 
In fact, the extraction of a contaminated spectrogram from the full dispersed image of a detector is not perfect, since the same wavelength may be at the beginning, middle or end of a pixel, depending on the dispersion direction.

Finally, it is essential to inter-calibrate the flux of  the four spectrograms in the four dispersion directions by multiplying them with relative  correction factors. The goal of this preprocessing is to bring all spectrograms to a common instrumental flux scale \citep{Calib}.

After applying these three preprocessings, we then define the total observation matrix ${\bf X}$ of an object of interest by concatenating its contaminated spectrograms in the four directions as follows

\begin{equation}
{\bf X}= \begin{bmatrix} 
{\bf X}^{(d_1)}\\ 
{\bf X}^{(d_2)}\\ 
{\bf X}^{(d_3)}\\
{\bf X}^{(d_4)}\\ 
\end{bmatrix}=	\begin{bmatrix} 
{\bf X}^{(0)}\\ 
{\bf X}^{(180)}\\ 
{\bf X}^{(184)}\\
{\bf X}^{(-4)}\\ 
\end{bmatrix}={\bf A} {\bf E},
\label{Mix_LI}
\end{equation}	
where 
${\bf A}=\left[ {\bf a}_s|{\bf A}_c\right] $  is the total mixing matrix of size 
$M \times N$, with $M=\sum_{i=1}^{4}M_i$, $N=(\sum_{i=1}^{4}N_i)-3$, 
${\bf a}_s$ is the mixing vector of the object of interest,
defined by 
\begin{equation}
{\bf a}_s=[{\bf a}_s^{{(1)}^T},{\bf a}_s^{{(2)}^T}, {\bf a}_s^{{(3)}^T}, {\bf a}_s^{{(4)}^T}]^T,
\end{equation} 
and ${\bf A}_c$ is the contaminant mixing matrix. Assuming the contaminants are different in each dispersion direction, this matrix can be written as
\begin{equation}
{\bf A}_c=\begin{bmatrix} 
{\bf A}_c^{(1)}&0&0&0\\ 
0&{\bf A}_c^{(2)}&0&0\\ 
0&0&{\bf A}_c^{(3)}&0\\ 
0&0&0&{\bf A}_c^{(4)}\\ 
\end{bmatrix}.
\label{MLI03}
\end{equation}
${\bf E}$ is the matrix of size $N \times K$ containing convolved version of the spectrum of the object of interest in its first row and convolved versions of the spectra of  contaminants in each of the four directions in the following rows. It should be noted that if all contaminants are considered,  the mixing matrix $\bf{A}$ may become underdetermined (i.e. $M < N$). In this case, we retain only the brightest contaminants in the four  dispersion directions, so that the mixing matrix $\bf{A}$ becomes (over)-determined (i.e. $M\geq N$).
In Sect. \ref{sec:LI method}, we will present three methods that can be used to estimate the spectrum of the object of interest by exploiting the observation matrix ${\bf X}$ and available information on the objects.

\subsection{Local convolutive mixing model}
\label{sec_MC}
We recall the expression of the spectrogram model, as presented in Eq. (\ref{modeli3}), for an object with index $j$ dispersed in the horizontal direction $x$:
\begin{equation}
t_j( x , y )=\int_{\Omega_\lambda}^{}w_j( x-D(\lambda) , y , \lambda)\;\mathrm{d}\lambda,
\label{model3}
\end{equation}
where $w_j( x , y , \lambda)=[f_j ( x, y )* h( x , y )] s_j (\lambda )$.
Note that this equation is valid for all dispersion directions $d_i\in{ 0^\circ, 180^\circ, 184^\circ, -4^\circ}$ after rotating the spectrograms for each direction according to their respective angles. 

Eq. (\ref{model3}) can be simplified by calculating its Fourier Transform (FT) in the spatial domain as follows 
\begin{equation}
\begin{aligned}[b]
T_j( f_x , f_y )&=\text {FT}\left\{t_j( x , y )\right\}=\text{FT} \left\{\int_{\Omega_\lambda}^{}w_j( x-D(\lambda) , y , \lambda)\;\mathrm{d}\lambda \right\}\\
&=\int_{\Omega_\lambda}^{}W_j( f_x, f_y , \lambda) \mathrm{e}^{-\mathrm{i}2\pi f_x D(\lambda)} \; \mathrm{d}\lambda\\
&=F_j( f_x, f_y)\, H( f_x, f_y) \int_{\Omega_\lambda}^{} s_j(\lambda)\mathrm{e}^{-\mathrm{i}2\pi f_x D(\lambda)}\;\mathrm{d}\lambda,
\end{aligned}
\label{model4}
\end{equation}
where $F_j(f_x,f_y)$ and $H(f_x,f_y)$ are the 2D Fourier transforms of $f_j(x,y)$ and $h(x,y)$, respectively, and the variables $f_x$ and $f_y$ correspond to the spatial frequency components in the Fourier domain.

Considering a linear dispersion $D(\lambda)=\alpha \lambda+\beta$ (where $\alpha$ and $\beta$ are two known coefficients), Eq. (\ref{model4}) becomes
\begin{equation}
\begin{aligned}[b]
T_j( f_x , f_y )&=G_j( f_x, f_y) \int_{\Omega_\lambda}^{} s_j(\lambda)\mathrm{e}^{-\mathrm{i}2\pi f_x \alpha\lambda}\;\mathrm{d}\lambda \, \mathrm{e}^{-\mathrm{i}2\pi f_x \beta}\\
&=G_j( f_x, f_y) \, S_j(\alpha f_x) \, \mathrm{e}^{-\mathrm{i}2\pi f_x \beta},
\end{aligned}
\label{model5}
\end{equation}
where $G_j( f_x, f_y)=F_j( f_x, f_y)\, H( f_x, f_y)$ and $S_j$ is the 1D Fourier transform of $s_j$.
Note that $S_j(\alpha f_x)$ does not depend on the vertical frequency coordinate $f_y$.

The observed value at a frequency indexed by $(f_x, f_y)$ in the dispersion direction $d_i$ is expressed as follows
\begin{equation}
\begin{split}
O^{{(d_i)}}( f_x , f_y )&=\sum_{j=1}^{N_i}T^{{(d_i)}}_j( f_x , f_y )\\
&=\sum_{j=1}^{N_i}G^{{(d_i)}}_j( f_x, f_y) \, S_j(\alpha f_x) \, \mathrm{e}^{-\mathrm{i}2\pi f_x \beta},
\end{split}
\label{model6}
\end{equation}
where $N_i$ is the number of objects in the dispersion direction $d_i$.
Using vector notations, Eq. (\ref{model6}) can be reformulated as a complex-valued linear instantaneous model at each horizontal frequency $f_x$:
\begin{equation}
{\bf X}^{{(d_i)}}( f_x )= {\bf A}^{(i)}(f_x) \, {\bf S}^{(i)}(f_x),\label{model7}
\end{equation}
where ${\bf X}^{{(d_i)}}(f_x)=[O^{{(d_i)}}(f_x,f_y(1)),\cdots,O^{{(d_i)}}(f_x,f_y(M_i))]^{\rm T}$ is a vector of size $M_i \times 1$ containing the observed data at the horizontal frequency $f_x$, with $M_i$ being the number of vertical frequencies in the cross-dispersion direction $y$. 
$M_i$ can vary from one dispersion direction to another, and its value depends on the vertical extent of the object of interest in that dispersion direction.
Since the sources can be freely rearranged, we will assume throughout  the rest of this article that the first source $S_1(f_x)$ is the source of interest, and that
${\bf S}^{(i)}( f_x)=[S_1(\alpha f_x),S^{(d_i)}_{2}(\alpha f_x),\cdots,S^{(d_i)}_{N_i}(\alpha f_x)]^{\rm T}$ is a vector of size $N_i \times 1$ containing compressed versions of the Fourier transforms of the source spectra. \\
${\bf A}^{(i)}(f_x)=\left[ {\bf a}_s^{(i)}(f_x)|{\bf A}_c^{(i)}(f_x)\right]\, \mathrm{e}^{-\mathrm{i}2\pi f_x \beta}$ is the mixing matrix of size $M_i \times N_i$, where $\left[ ...|... \right]$ is the concatenation of 2 matrices, ${\bf a}_s^{(i)}(f_x)=[G_1^{(d_i)}(f_x,f_y(1)),\cdots,G_1^{(d_i)}(f_x,f_y(M_i))]^{\rm T}$ is the mixing vector of the object of interest at each horizontal frequency $f_x$, and
${\bf A}_c^{(i)}(f_x)$ is the contaminant mixing matrix of size $M_i \times (N_i-1)$, defined by 
\begin{equation}\label{mixing_matrix0}
{\bf A}_c^{(i)}(f_x)= \begin{bmatrix} 
G^{(d_i)}_2(f_x,f_y(1)) & \ldots & G^{(d_i)}_{N_i}(f_x,f_y(1)) \\
\vdots    & \ddots  & \vdots \\
G^{(d_i)}_2(f_x,f_y(M_i)) & \cdots& G^{(d_i)}_{N_i}(f_x,f_y(M_i))
\end{bmatrix}.
\end{equation}

We have four observations for each object of interest, corresponding to the four dispersion directions $d_i\in\{ 0^\circ, 180^\circ, 184^\circ, -4^\circ\}$. Since the spectrum of the object of interest is the same regardless of the dispersion direction, it is possible to combine these four observations to improve the estimation of its spectrum. 
We define the total observation vector ${\bf X}(f_x)$ corresponding to the object to be decontaminated at each horizontal frequency $f_x$ by merging the Fourier transforms of its contaminated spectrograms in the four directions as follows
\begin{equation}
{\bf X}( f_x )= \begin{bmatrix} 
{\bf X}^{(0)}( f_x )\\ 
{\bf X}^{(180)}( f_x )\\ 
{\bf X}^{(184)}( f_x )\\
{\bf X}^{(-4)}( f_x )\\ 
\end{bmatrix}={\bf A}( f_x ) {\bf S}( f_x ),
\label{Mix}
\end{equation}	
where ${\bf A}( f_x )$ is the total mixing matrix of size 
$M \times N$, with $M=\sum_{i=1}^{4}M_i$, $N=(\sum_{i=1}^{4}N_i)-3$, defined by
\begin{equation}
\begin{split}
{\bf A}( f_x )&= \mathrm{e}^{-\mathrm{i}2\pi f_x \beta} \times\\
&\begin{bmatrix}
{\bf a}_s^{(1)}( f_x )&{\bf A}_c^{(1)}( f_x )&0&0&0\\ 
{\bf a}_s^{(2)}( f_x )&0&{\bf A}_c^{(2)}( f_x )&0&0\\ 
{\bf a}_s^{(3)}( f_x )&0&0&{\bf A}_c^{(3)}( f_x )&0\\ 
{\bf a}_s^{(4)}( f_x )&0&0&0&{\bf A}_c^{(4)}( f_x )\\ 
\end{bmatrix}.
\label{MLI3}
\end{split}
\end{equation}

Finally, ${\bf S}( f_x)$ is a vector of size $N \times 1$ containing the Fourier transform of the spectrum of the object of interest at the horizontal frequency $f_x$ in its first row, followed by the Fourier transforms of the contaminant spectra at the horizontal frequency $f_x$ in each of the four directions in the following rows. This vector is defined as
\begin{equation}\label{vec_s}
\scalebox{0.9}{$
\begin{split}
{\bf S}( f_x)=&[S_1(\alpha f_x),S^{(0)}_{2}(\alpha f_x),\cdots,S^{(0)}_{N_1}(\alpha f_x),S^{(180)}_{2}(\alpha f_x),\cdots,S^{(180)}_{N_2}(\alpha f_x),\\
&S^{(184)}_{2}(\alpha f_x),\cdots,S^{(184)}_{N_3}(\alpha f_x),S^{(-4)}_{2}(\alpha f_x),\cdots,S^{(-4)}_{N_4}(\alpha f_x)]^{\rm T}.
\end{split}$}
\end{equation}	

It should be noted that prior to the combination of the spectrograms from the four directions, the three preprocessings (rotations, resampling and calibration) introduced in Sect. \ref{sec_MLI} must be applied to the spectrograms of the object of interest in the spatial domain.

\section{Methods}
\label{proposed}
\subsection{Local instantaneous methods}
\label{sec:LI method}

	The local instantaneous approach is based on the linear instantaneous mixing model introduced in Sect. \ref{sec_MLI} and exploits the direct images of the objects to estimate the spectrum of an object of interest.
This approach provides the vectors ${\bf e}_j$ defined in Eq. (\ref{mod loc4}), which represent estimates of the spectra ${\bf s}_j$ after convolution by a function depending on the object's profile in the direction of dispersion. The aim here is to separate the spectra, not deconvolve them.

\subsubsection{Mixing matrix estimation}
\label{sec:first method}
The first step, common to all the proposed local instantaneous methods, consists in estimating the mixing matrix ${\bf A}$, using the direct image of the object of interest as well as those of its contaminants. As previously mentioned in   Sect. \ref{introduction}, \Euclid is also equipped with a photometer that provides three direct images in the three spectral bands $Y$, $J$ and $H$ of all astronomical objects in the field of view \citep{Laureijs11}.
However, in this work, only direct images of the $J$ band were available and will be used to estimate the mixing matrix. 
The direct image, denoted $p^{{(d_i)}}_j(x,y)$, is defined for an object of index $j$ in the dispersion direction $d_i$ as the integral of the light intensity of the observed object in Eq. (\ref{model2}) over the spectral domain:
\begin{equation}
p^{{(d_i)}}_j( x , y )=\int_{\Omega_{J}}^{}w^{{(d_i)}}_j( x , y , \lambda)\;\mathrm{d}\lambda = I_j^{{(d_i)}}( x, y ) \int_{\Omega_{J}}^{}s_j (\lambda )\;\mathrm{d}\lambda,
\label{Im0}
\end{equation}
where $\Omega_{J}$ represents the wavelength range covered by the $J$ band and $I_j( x, y )= f_j(x, y) * h(x,y)$ is the profile of the object with index $j$ convolved to the instrument's PSF.

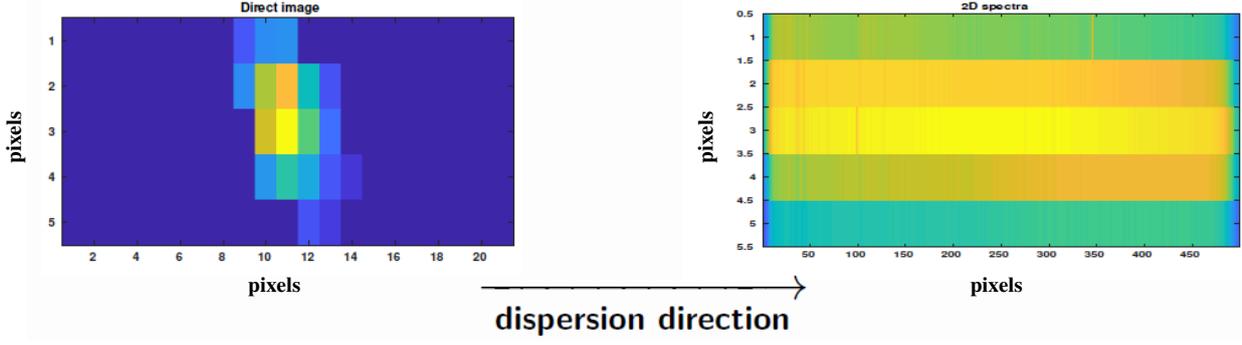
\begin{figure*}[tp!]
	\centering	
		\begin{tikzpicture}	
	\node  at (-0.0,0) {\includegraphics[width=16.5cm, keepaspectratio]{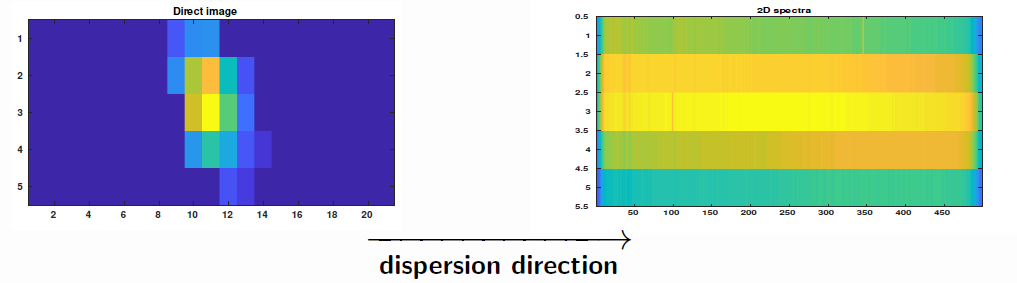}};	
	\node [rotate=90]  at (-8.4,0.4) {\tiny{\bf pixels}}; 	 
	\node  at (-5.0,-1.58) {\tiny{\bf pixels}};	 	 
		\node [rotate=90]  at (0.7,0.4) {\tiny{\bf pixels}}; 	 
	\node  at (4.5,-1.58) {\tiny{\bf pixels}};	 	 	 
	\end{tikzpicture}
	\caption{Flux distribution of the direct image and the uncontaminated spectrogram of an object.}
	\label{flux}
\end{figure*}

As illustrated in Fig. \ref{flux}, the flux of the direct image is dispersed by a grism to produce its spectrogram, also known as a 2D spectrum.
Assuming that the light conserves its flux after dispersion, the sum of the fluxes on the pixels of each row of the direct image is equal to the sum of the fluxes on the pixels of each row of the image dispersed by the grism (assuming that the direct image and the dispersed image are observed over the same wavelength range).
In this case, the flux distribution of the direct image in the direction orthogonal to the dispersion direction (called the cross-dispersion direction) is identical to that of the flux of the uncontaminated spectrogram of an object of interest. Consequently, the mixing coefficients $a_{j}^{(d_i)}(m)$ associated with this object (whether the target object or a contaminant) can be estimated by calculating the sum of the pixel values along the different rows of its direct image.
However, since the grism and the direct image of the $J$ band do not cover exactly the same spectral band, the sum of each row does not  precisely yield $a_{j}^{(d_i)}(m)$, but rather an estimate of $a_{j}^{(d_i)}(m)$. However, if we model this estimation error by a multiplicative coefficient $k$, the value of $k$ is constant over all the rows of the image. This can be thought of as the scale factor indeterminacy in a classical source separation problem\footnote{By simultaneously using the three direct NISP images in the $Y$, $J$ and $H$ bands, we can reduce this estimation error. However, during our work, we only had access to the image in the $J$ band.}.

To estimate the coefficients $a_j^{(d_i)}(m)$, we first oversample the direct image of each object in the cross-dispersion direction, then shift this image in the cross-dispersion direction in order to recenter it with respect to the position of the object of interest, and finally undersample it by the same sampling rate. This allows us to correct the offset between the direct image and the spectrogram of this object.
Secondly, as the optics used in \Euclid's photometer and spectrometer are different, their PSFs are not the same.
It is therefore necessary to correct the photometric PSF by convolving the direct image of each object with the differential PSF between the photometer and the spectrometer of \Euclid, which we assume to be independent of wavelength.  This pre-processing is essential to ensure that the flux distribution of the direct image in the  cross-dispersion direction is identical to that of the flux of the uncontaminated spectrogram of an object of interest.

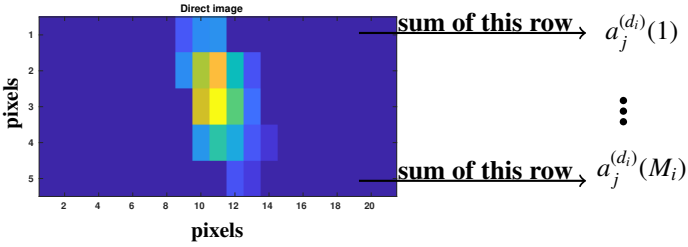
\begin{figure}[h]
	\hspace*{-0.4cm}
	\centering
	\begin{tikzpicture}[scale=0.49]	
	\node  at (-0.5,2.2) {\includegraphics[width=5cm, keepaspectratio]{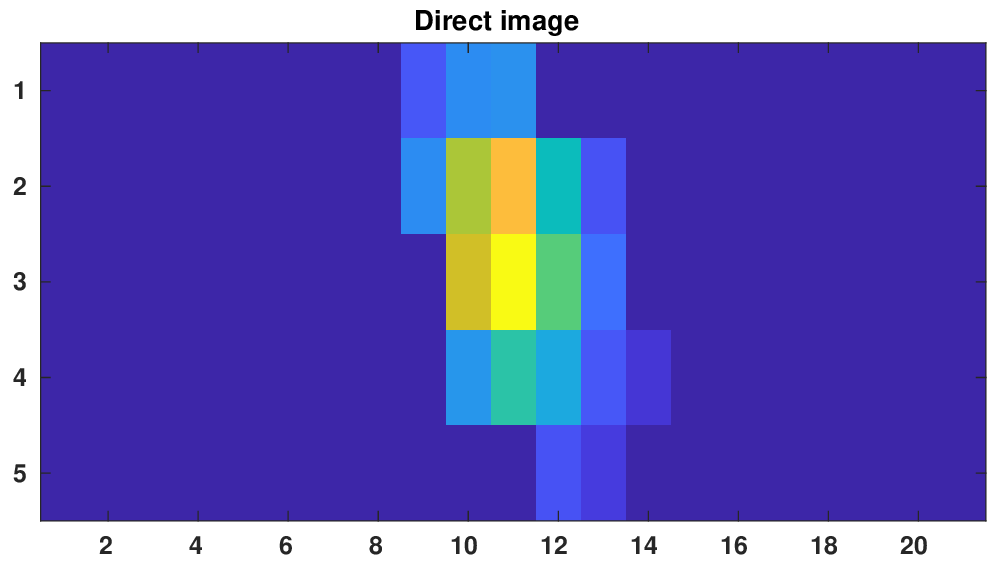}};
	\draw [black, thick,->]  (3.4,4.3) -- (9.5,4.3);
	\draw [black, thick,->]  (3.4,0.3) -- (9.5,0.3);
	\node  at (6.8,4.6) {{\bf sum of this row}};
	\node  at (6.8,0.6) {{\bf sum of this row}};
	\node  at (11,4.3) {{$a_{j}^{(d_i)}(1)$}};
	\node  at (11,0.6) {{$a_{j}^{(d_i)}(M_i)$}};
	\node  at (10.5,2.4) {\Huge{$\vdots$}};		
	\node [rotate=90]  at (-5.9,2.4) {\tiny{\bf pixels}}; 	 
	\node  at (-0.4,-1.08) {\tiny{\bf pixels}};	 	 	 	 
	\end{tikzpicture}
	\caption{Mixing coefficient estimation from direct images.}
	\label{fig:Mixing}
\end{figure}

After these preprocessings, we select the $M_i\times R$ pixel values of the image centred on the position of the object of interest, where $R$ is the number of columns in the direct image and $M_i$ is the number of rows in  the cross-dispersion direction of the spectrogram. Then the sum of the pixel values of the {\it m}-th row gives the estimated  value of the mixing coefficient $a_{j}^{(d_i)}(m)$ for that object. This process is illustrated in  Fig. \ref{fig:Mixing}. 	The mixing matrix ${\bf A}$ is then constructed using the estimated coefficients.

\subsubsection{Source matrix estimation}	
A first solution to estimate the source matrix ${\bf E}$ is to minimize the following criterion, where $||.||_2$ is the Frobenius norm 
\begin{equation}
J_1=||{\bf X}-{\bf A}{\bf E}||_{2}^2,
\label{lsq1}
\end{equation}
which leads to Least SQuares (LSQ) estimation:
\begin{equation}
{\bf E}=	({\bf A}^{T}{\bf A})^{-1}{\bf A}^{T}{\bf X}.
\label{lsq2}
\end{equation}
A second solution for estimating ${\bf E}$ is to use the Least SQuares with Non-negativity constraint (LSQN) algorithm presented in \citep{Lawson1974}. Since the spectra to be estimated are by definition non-negative, this method takes this non-negativity into account by minimizing the criterion:
\begin{equation}
J_2=||{\bf X}-{\bf A}{\bf E}||_{2}^2\,\,\, \text{s.t.}\,\,\, \	{\bf E}\geq 0
\label{nnlsq}
\end{equation}
where ${\bf E}\geq 0$ means that all values of the matrix ${\bf E}$ are non-negative.
Finally, the convolved version of the spectrum of the object of interest, denoted ${\hat{\bf e}}_{1}$, is the first row of the estimated matrix ${\hat{\bf E}}$.

Unlike the first two methods, which require both the direct image of the object of interest and all its contaminants, our third method requires only the image of the object of interest to estimate its spectrum.
In fact, this image allows us to estimate the mixing coefficients $a_{1}^{(d_i)}(m)$ of the object of interest in all dispersion directions, which will allow us to estimate the mixing vector ${\bf a}_s$ of this object by the same procedure as that described in Sect. \ref{sec:first method}. This vector is then used by a beamforming method to estimate the vector ${\hat{\bf e}}_{1}$. 
The goal of the beamformer  is to extract a desired signal (spectrum of the object of interest) while attenuating interferences (spectra of contaminants) and noise \citep{MVDR,beamforming}.
In this paper, we used the beamformer known as the Minimum Power Distortionless
Response (MPDR). This beamformer aims to estimate an optimal filter denoted ${\bf w}_{\text{MPDR}}$, whose output minimizes the total power under the constraint ${\bf w}^H{\bf a}_s=1$. Assuming that the spectrum of the object of interest is uncorrelated with the spectra of the contaminants, the desired filter is the solution to the following minimization problem
\begin{equation}
{\bf w}_{\small \text{MPDR}}=\operatorname*{argmin}_{{\bf w}}\:\left\{{\bf w}^H{\bf R}_{\bf X}{\bf w}\right\} \,\,\, \text{s.t.}\,\,\, \	{\bf w}^H{\bf a}_s=1,
\label{beam1}
\end{equation}
which leads to the following beamforming coefficients \citep{beamforming}
\begin{equation}
{\bf w}_{\small \text{MPDR}}=\frac{{\bf R}_{\bf X}^{-1}{\bf a}_s}{{\bf a}_s^H{\bf R}_{\bf X}^{-1}{\bf a}_s},
\label{beam20}
\end{equation}
where ${\bf R}_{\bf X}= ({\bf X}-\text{mean}({\bf X}))({\bf X}-\text{mean}({\bf X}))^H/K $ is the covariance matrix of the observations, $\text{mean}({\bf X})$ is the mean over all rows of the matrix ${\bf X}$, and $K$ is the number of spectral bands. 

Finally, the vector ${\hat{\bf e}}_{1}$ is estimated as follows
\begin{equation}
\hat{{\bf e}}_{1} = {\bf w}_{\small \text{MPDR}}^{H}{\bf X}.
\label{images00}
\end{equation}

Although MPDR beamforming can be applied directly to the total observation matrix ${\bf X}$ to estimate the spectrum of the object of interest, we propose here to split the total observation matrix ${\bf X}$ into $L$ distinct parts. Each part, denoted ${\bf X}_l$, includes all rows of the matrix ${\bf X}$ and a specific set of columns of this matrix. We then decontaminate each part separately. The goal of this approach is to minimize the risk of encountering underdetermined cases that MPDR beamforming cannot effectively handle. This is because each contaminating object generally contaminates only a portion of the spectrogram of the object of interest. Consequently, the number of contaminants for each of the $L$ parts is less than the number of contaminants for the entire spectrum.
To implement this method, we generate $L$ beamformers, each given by 
\begin{equation}
{\bf w}_{\small \text{MPDR}}^{(l)}=\frac{{\bf R}_{{\bf X}_l}^{-1}{\bf a}_s}{{\bf a}_s^H{\bf R}_{{\bf X}_l}^{-1}{\bf a}_s},\hspace*{0.3cm} l\in \left[1,L\right], 
\end{equation}
where ${\bf R}_{{\bf X}_l}$ is the covariance matrix associated with the $l-$th part of the observation matrix ${\bf X}$.
Finally, the outputs $ \hat{{\bf e}}_{1}^{(l)} = {\bf w}_{\small \text{MPDR}}^{(l)H}{\bf X}_l$ of each beamformer are concatenated to obtain the final estimate of the vector ${\hat{\bf e}}_{1}$ as follows
\begin{equation}
\hat{{\bf e}}_{1}=[\hat{{\bf e}}_{1}^{(1)},\hat{{\bf e}}_{1}^{(2)},...,\hat{{\bf e}}_{1}^{(L)}].
\label{images000}
\end{equation}
Note that this method also attenuates unmodeled interferences, such as the spectra of undetected objects or hot pixels, which should lead to better results under these conditions. 

Knowing that each column of the  mixing matrix $\bf{A}$ is estimated up to a scaling coefficient, the three local instantaneous methods proposed in this section estimate  ${\bf e}_{1}$ up to a scaling factor.

\subsection{Local convolutive method}
\label{conv_meth}
Unlike the approach presented in Sect. \ref{sec:LI method}, which is based on an approximate linear instantaneous model, the local convolutive approach is based on a more realistic convolutive mixing model introduced in Sect. \ref{sec_MC}. This model allows simultaneous decontamination and deconvolution of  spectra.

\subsubsection{Mixing matrix estimation}
We recall the definition of the direct image $p^{{(d_i)}}_j(x,y)$, as given in Eq. (\ref{Im0}), for an astronomical object with index $j$ in the dispersion direction $d_i$:
\begin{equation}
\begin{split}
p^{{(d_i)}}_j( x , y )&=\int_{\Omega_{J}}^{}w^{{(d_i)}}_j( x , y , \lambda)\;\mathrm{d}\lambda,\\
&=f^{{(d_i)}}_j ( x, y )* h^{{(d_i)}}( x , y )\cdot \int_{\Omega_{J}}^{}s_j (\lambda )\;\mathrm{d}\lambda.
\end{split}
\label{Im1}
\end{equation}
By computing the 2D Fourier transform in the spatial domain and assuming that the direct image $p^{{(d_i)}}_j( x, y )$ is centered with respect to the position of the object of interest in the two spatial dimensions $x$ and $y$, we obtain

\begin{equation}
\begin{split}
I^{{(d_i)}}_j( f_x , f_y )&=G^{{(d_i)}}_j( f_x, f_y)\,\int_{\Omega_{J}}^{} s_j(\lambda)\;\mathrm{d}\lambda,\\
&=G^{{(d_i)}}_j( f_x, f_y)\, C_j,
\end{split}
\label{Im2}
\end{equation}
where $I^{{(d_i)}}_j$ is the 2D Fourier transform of $p^{{(d_i)}}_j$ and $C_j=\int_{\Omega_{J}}^{} s_j(\lambda)\;\mathrm{d}\lambda$ is a constant independent of the spatial frequencies $f_x$ and $f_y$.
Consequently, the mixing coefficients $G^{{(d_i)}}_j(f_x,f_y)$ in the Fourier domain corresponding to a given object of index $j$ (target or contaminant), can be estimated up to factors  $C_j$ using direct images for each dispersion direction. This allows us to estimate the total mixing matrix ${\bf A}(f_x)$ defined in (\ref{MLI3}). 

However, before estimating the mixing coefficients $G^{{(d_i)}}_j(f_x,f_y)$ using the direct images, some pre-processing is required. Firstly, it is essential to align the direct image of each object with the spectrogram of the object of interest in both spatial dimensions, $x$ and $y$, which is a necessary condition for the validity of Eq. (\ref{Im2}).
Secondly, it is  necessary to correct the photometric PSF by convolving it with the differential PSF between \Euclid's photometer and spectrometer.
Thirdly, it is necessary to normalize the Fourier transform of each image to avoid the introduction of scaling factors.

Once these preprocessings are completed, the mixing vectors of the object of interest ${\bf a}_s^{(i)}(f_x)$ can be estimated using the direct images $I^{(d_i)}_s(f_x, f_y)$ of the object of interest, and the contaminant mixing matrices ${\bf A}_c^{(i)}(f_x)$ can be estimated using the direct images of the contaminants $I^{(d_i)}_j(f_x, f_y)$.
The total mixing matrix ${\bf A}(f_x)$ defined in (\ref{MLI3}) is then constructed using the calculated matrices ${\bf A}_c^{(i)}(f_x)$ and the vectors ${\bf a}_s^{(i)}(f_x)$.

The procedure for estimating the mixing matrix at each horizontal frequency $f_x$ is presented in Algorithm \ref{matrice_alg}.
\begin{algorithm}[h]
	\caption{Mixing matrix estimation at each horizontal frequency $f_x$ in the local convolutive approach}\label{matrice_alg}
	\KwIn{Direct images $p^{{(d_i)}}_j(x,y)$ of all objects in all directions  $d_i\in\{ 0^\circ, 180^\circ, 184^\circ, -4^\circ\}$.}
	\KwResult{Mixing matrix ${\bf A}( f_x )$}
	\For{$d_i=\{ 0^\circ, 180^\circ, 184^\circ, -4^\circ\}$}{
		Align the image $p^{{(d_i)}}_1(x,y)$ with the spectrogram of the object of interest\\
		Convolve the image $p^{{(d_i)}}_1(x,y)$ with the differential PSF\\
		Compute the 2D Fourier transform in the spatial domain of the image $p^{{(d_i)}}_1(x,y)$\\
		Normalize the Fourier transform of the image $I^{{(d_i)}}_1( f_x , f_y )$\\
		Construct the mixing vector of the object of interest ${\bf a}_s^{(i)}(f_x)$\\
		\For{$j=2$ to $N_i$}{
			Align the image $p^{{(d_i)}}_j(x,y)$ with the spectrogram of the object of interest\\
			Convolve the image $p^{{(d_i)}}_j(x,y)$ with the differential PSF\\
			Compute the 2D Fourier transform in the spatial domain of the image $p^{{(d_i)}}_j(x,y)$\\
			Normalize the Fourier transform of each image $I^{{(d_i)}}_j( f_x , f_y )$\\
			
		}		{\bf construct}{
			${\bf A}_c^{(i)}(f_x)$}
	}
	\Return{${\bf A}( f_x )$}
\end{algorithm}
\subsubsection{Source matrix estimation}
After constructing the mixing matrix ${\bf A}(f_x)$ for each horizontal frequency $f_x$, the next step is to estimate the source matrix ${\bf S}(f_x)$.
However, the mixing matrix ${\bf A}$ is often underdetermined (i.e., $M<N$), making it non-invertible. To solve this problem, we have assumed the following two realistic assumptions
\begin{enumerate}
	\item At most $N_c$ ($N_c\leq M-1$) contaminants are active at each horizontal frequency $f_x$.
	
	\item 	The source of interest is active at all horizontal frequencies $f_x$.
\end{enumerate}

Next, the procedure for detecting the indices of the $N_c$ contaminants active in each horizontal frequency $f_x$ is presented in Algorithm \ref{sources_alg}, which can be seen as a modified version of the Orthogonal Matching Pursuit (OMP) algorithm \citep{OMP,OMP2}. In Algorithm \ref{sources_alg}, ${\rm H}$ denotes the Hermitian transpose, $\boldsymbol{{ind}}(f_x)$ is the vector of indices related to the object of interest and the dominant contaminants, ${\bf A}( : ,\boldsymbol{{ind}}(f_x))$ represents a submatrix of ${\bf A}(f_x)$ with all the rows and only the columns present in the vector $\boldsymbol{{ind}}(f_x)$ at the horizontal frequency $f_x$, and $pinv$ is the pseudo-inverse of a matrix, defined by $pinv({\bf M})=({\bf M}^H{\bf M})^{-1}{\bf M}^H$.
The algorithm returns an index vector $\boldsymbol{{ind}}(f_x)$ indicating which contaminants are detected in each horizontal frequency $f_x$. It works by  initializing a residual $\boldsymbol{r}$, then iterates to find the contaminant indices one by one by maximizing  the correlation between the residual $\boldsymbol{r}$ and the different columns of the matrix ${\bf A}(f_x)$. Once a contaminant is detected, it is removed from the residue, and the process continues until all $N_c$ contaminants have been identified.

\begin{algorithm}[]
	\caption{Detection of active contaminants for each horizontal frequency $f_x$}\label{sources_alg}
	\KwIn{Mixing matrix ${\bf A}(f_x)$, observation vector ${\bf X}( f_x )$ and the assumed number of active contaminants $N_c$.}
	\KwResult{Index vector $\boldsymbol{{ind}}(f_x)$}
	\textbf{Initialization:} $\boldsymbol{{ind}}(f_x)=[1]$, $\boldsymbol{r}={\bf X}( f_x )-{\bf A}(:,\boldsymbol{{ind}}(f_x))(pinv({\bf A}(:,\boldsymbol{{ind}}(f_x))){\bf X}( f_x ))$ \\
	\For{$i=1$ to $N_c$}{
		Find the index $i$ that maximizes $|{\bf A}^{\rm H}(:,i)\boldsymbol{r}|$\\
		Set $\boldsymbol{{ind}}(f_x)=[\boldsymbol{{ind}}(f_x), i]$\\
		Set $\boldsymbol{r}={\bf X}( f_x )-{\bf A}(:,\boldsymbol{{ind}}(f_x))(pinv({\bf A}(:,\boldsymbol{{ind}}(f_x))){\bf X}( f_x ))$
	}
	\Return{$\boldsymbol{{ind}}(f_x)$}
\end{algorithm}

Once the vector $\boldsymbol{{ind}}(f_x)$ is estimated, we propose to use a beamforming technique to estimate the Fourier transform of the spectrum of the object of interest. In our method, we used the beamformer called Linearly Constrained Minimum Power (LCMP) \citep{LCMP}. This beamformer aims to estimate an optimal filter denoted by ${\bf w}_{\small \text{LCMP}}(f_x)$ at each horizontal frequency $f_x$, whose output minimizes the total power under the constraint ${\bf w}^{\rm H}(f_x){\bf A}(:,\boldsymbol{{ind}}(f_x))=[1,0,...,0]$. The filter we are looking for is the solution to the following minimization problem  \citep{LCMP} 
\begin{equation}
\begin{split}
{\bf w}_{\small \text{LCMP}} (f_x)=&\operatorname*{argmin}_{{\bf w}(f_x)}\:\left\{{\bf w}^{\rm H} (f_x){\bf R}_{\bf X}{\bf w}(f_x)\right\}\\
&\text{s.t.}\,\,\, \ {\bf w }^{\rm H} (f_x){\bf A}(:,\boldsymbol{{ind}}(f_x))=[1, 0, \cdots, 0],
\end{split}
\label{beam2}
\end{equation}
which leads to the following beamforming coefficients
\begin{equation}
{\bf w}_{\small \text{LCMP}}(f_x)=\frac{{\bf R}_{\bf X}^{-1}{\bf A}(:,\boldsymbol{{ind}}(f_x))}{{\bf A}^{\rm H}(:,\boldsymbol{{ind}}(f_x)){\bf R}_{\bf X}^{-1}{\bf A}(:,\boldsymbol{{ind}}(f_x))}\, {\bf g},
\label{beam003}
\end{equation}
where ${\bf R}_{\bf X}= ({\bf X}-\text{mean}({\bf X}))({\bf X}-\text{mean}({\bf X}))^H/N_f $ is an estimate of the covariance matrix of the observations, ${\bf X}=[{\bf X}(1) \cdots, {\bf X}(N_f)]$ is a matrix combining the observation vectors for all frequencies, $N_f$ is the number of horizontal frequencies, and ${\bf g}=[1, 0, \cdots, 0]^{\rm T}$. 

Next, a compressed version of the Fourier transform of the spectrum of the object of interest ${\bf{S}}_{1}$ is estimated as follows\footnote{Note that the relations (\ref{beam003}) and (\ref{images12222}) are applied for each horizontal frequency $f_x$.}
\begin{equation}
{{S}}_{1}(f_x)= {\bf{w}}^{\rm H}_{\small \text{LCMP}} (f_x)	{{\bf X}(f_x)}.
\label{images12222}
\end{equation}

Finally, the spectrum of the object of interest is obtained by applying the Inverse  Fourier Transform (IFT) to the signals ${{\bf S}}_{1}$ as follows
\begin{equation}
{s}_1(\lambda)=\text{IFT}\{{{S}}_{1}(f_x)\}.
\label{images1222000}
\end{equation}	

For clarity, we will use the acronyms {\it LI-LSQ} (Linear Instantaneous LSQ), {\it LI-LSQN} (Linear Instantaneous LSQN), {\it LI-MPDR} (Linear Instantaneous MPDR) and {\it LC-LCMP} (Linear Convolutive LCMP) to refer to the different methods proposed in this section.

\section{Test results}
\label{test}

In this section, we will evaluate the decontamination performance of the methods proposed in this paper through four different experiments.
The first three experiments aim to evaluate the decontamination performance on three different scenarios, each representing a different level of contamination.    The fourth experiment evaluates these performances on a set of 100 different objects.
These evaluations were performed using realistic, noisy \Euclid-like simulations from the SC8 $R_{11}$ release, provided by the Euclid Consortium.
It is important to note that the observed data are also affected by high levels of noise introduced by the acquisition instruments.

\subsection{Performance measurement criteria}
To measure decontamination performance, we use three main criteria. The first criterion, called $\text{SIR}_\text{imp}$, evaluates the improvement in the Signal to Interference Ratio (SIR) before and after decontamination. It is defined as the difference between $\text{SIR}_\text{out}$ and $\text{SIR}_\text{in}$:
\begin{equation}
\text{SIR}_\text{imp}={\text{SIR}_\text{out}}-{\text{SIR}_\text{in}},
\end{equation}
where
\begin{equation}
\begin{cases}
{\text{ SIR}_\text{in}}=\frac{1}{4} \sum_{d_i= \{0, 180, 184, -4\}}10\;\mathrm{log_{10}}( \frac{||{\bf S}_s||_{2}}{||{\bf S}_{\bf x}^{(d_i)}-{\bf S}_s||_{2}}),\\
{\text{ SIR}_\text{out}}=10\;\mathrm{log_{10}}( \frac{||{\bf S}_s||_{2}}{||\hat{{\bf S}}_s-{\bf S}_s||_{2}}),
\label{SIR_in}
\end{cases}
\end{equation}
${\bf S}_{\bf x}^{(d_i)}$ is the 1D spectrum of the observed data (obtained by averaging all the rows of the observed spectrogram) in the $d_i$ direction, ${\bf S}_s$ is the true noiseless spectrum of the object of interest, and $\hat{\bf S}_s$ is its estimate obtained using a decontamination method. In (\ref{SIR_in}), all spectra are centered and $\hat{{\bf S}}_s$ and ${\bf S}_{\bf x}^{(d_i)}$ are normalized to have the same variance as ${\bf S}_s$.

The second criterion called  Root-Mean-Square Error ($\text{RMSE}$)  is defined as follows

\begin{equation}
{\text{ RMSE}}=\sqrt{\frac{\sum_{i=1}^{K}\left( \hat{S}_s(\lambda_i)-{S}_{s}(\lambda_i)\right)^2}{K}}.
\end{equation}

The third criterion, called the Local Root-Mean-Square Error (LRMSE), measures the estimation error within a small region centered around the true position of the main emission line. It is defined as:

\begin{equation}
{\text{LRMSE}}=\sqrt{\frac{\sum_{i\in \mathcal{L}}\left( \hat{S}_s(\lambda_i)-{S}_{s}(\lambda_i)\right)^2}{K_L}},
\end{equation}
where $\mathcal{L}$ denotes a set of $K_L$ wavelength indices corresponding to a small region around the true position of the main emission line of the object of interest. In all tests, $K_L$ is set to 20.
\\

It is important to note that $\text{SIR}_\text{imp}$ evaluates the improvement provided by the methods before and after decontamination in terms of interference reduction (mainly contaminant spectra), whereas $\text{RMSE}$ simply evaluates the estimation error after decontamination, 
and  $\text{LRMSE}$ specifically focuses on the estimation error around the main emission line.
 Better performance is indicated by a higher value of $\text{SIR}_\text{imp}$ and lower values of $\text{RMSE}$ and $\text{LRMSE}$.

\subsection{Simple scenario without hot pixels}
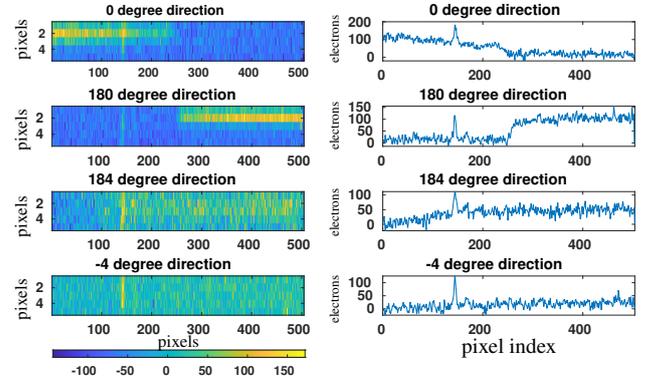
\begin{figure}[h!]
	\centering
			\hspace*{-0.3cm}
				\begin{tikzpicture}[scale=0.5]	
			\node  at (-0.5,2.2) {\includegraphics[angle=0,width=1.1\hsize]{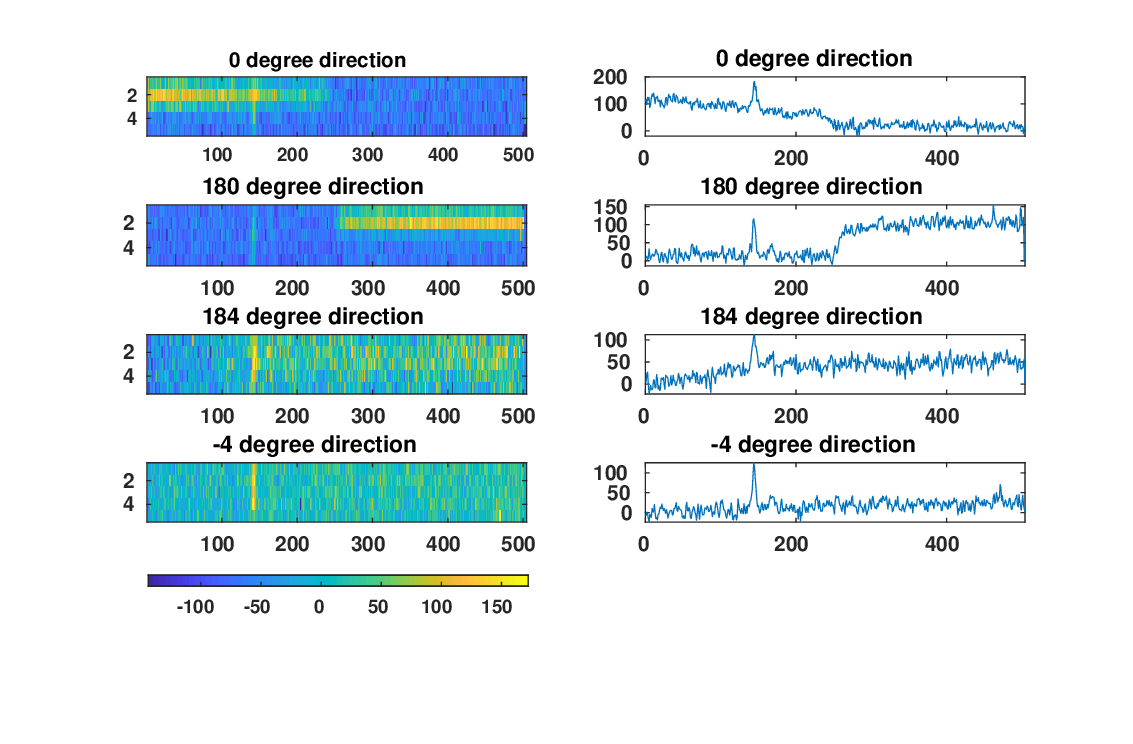}};
			\node  at (4.2,-1.3) {\tiny{pixel index}};
			\node [rotate=90]  at (-0.4,6.9) {\scalebox{.5}{electrons}}; 
			\node [rotate=90]  at (-0.4,4.6) {\scalebox{.5}{electrons}};
			\node [rotate=90]  at (-0.4,2.3) {\scalebox{.5}{electrons}}; 
			\node [rotate=90]  at (-0.4,-0.0) {\scalebox{.5}{electrons}}; 
			\node [rotate=90]  at (-8.6,6.9) {\scalebox{.65}{pixels}}; 
			\node [rotate=90]  at (-8.6,4.5) {\scalebox{.65}{pixels}};
			\node [rotate=90]  at (-8.6,2.2) {\scalebox{.65}{pixels}}; 
			\node [rotate=90]  at (-8.6,-0.0) {\scalebox{.65}{pixels}}; 	 
				\node  at (-4.5,-1.15) {\scalebox{.65}{pixels}};	 	 	 	 	 
			\end{tikzpicture}
		\vspace*{-1.5cm}
	\caption{Observed spectrograms (left) and corresponding 1D spectra (right) for the first scenario.}
	\label{observations}
\end{figure}

In our first experiment, we chose a simple scenario where the spectrogram of an object of interest (a star-forming  galaxy with a redshift of $1.12$) is slightly contaminated in all four dispersion directions. The observed spectrograms and corresponding 1D spectra\footnote{ The observed 1D spectrum is defined here as the average of all rows in the spectrogram. } for this object are shown in Fig. \ref{observations}.

\begin{figure}[h!]
	\centering
	\hspace*{-0.75cm}
	\includegraphics[width=10.25cm, keepaspectratio]{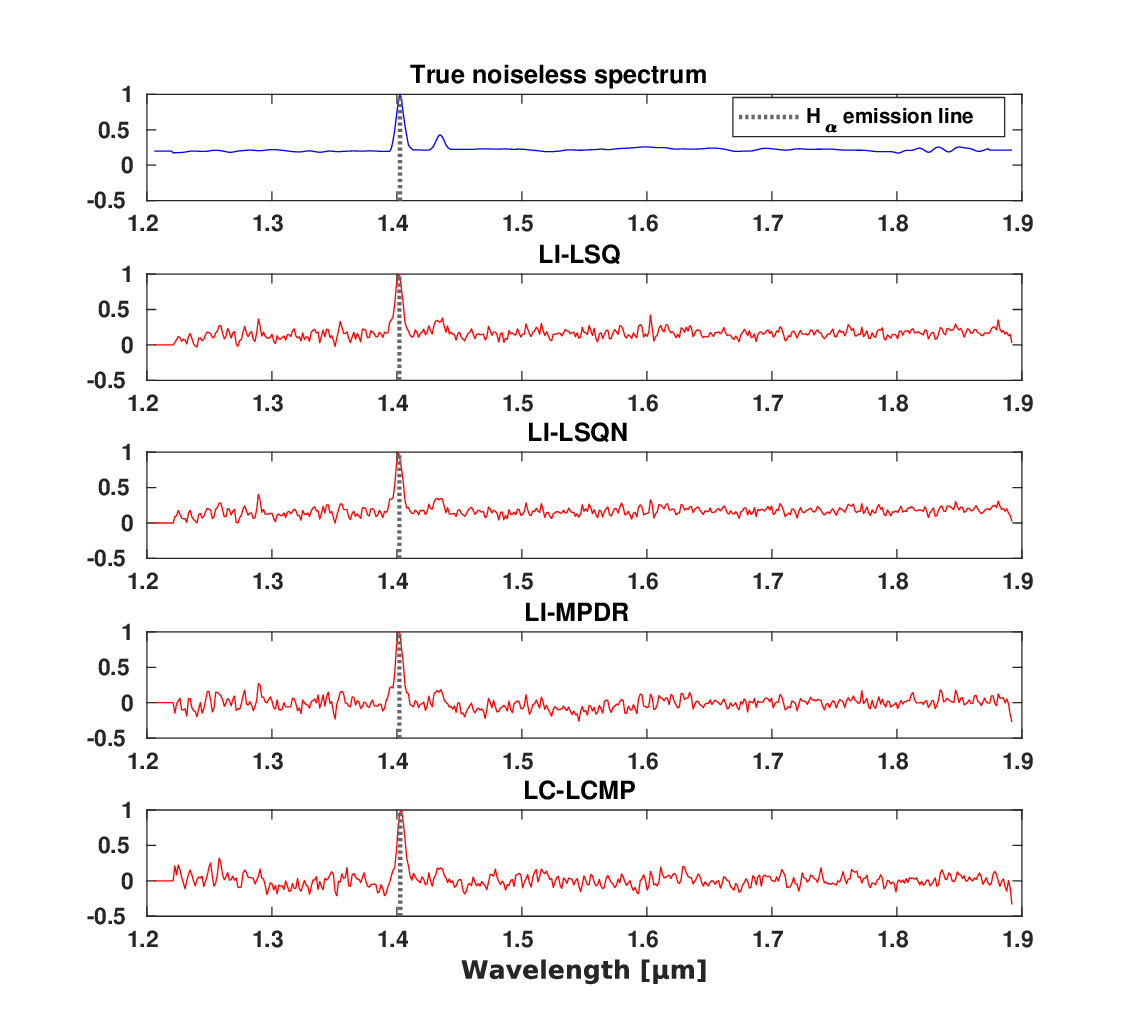}
	\vspace*{-0.5cm}
	\caption{True noiseless spectrum of the object of interest (in blue) and its estimates obtained using the proposed methods (in red) for the first scenario.}
\label{res_ex1}
\end{figure}

The true noiseless spectrum of the object of interest and its estimates using our methods {\it LI-LSQ}, {\it LI-LSQN}, {\it LI-MPDR} and {\it LC-LCMP} after normalization by their maximum are shown in Fig. \ref{res_ex1}. 
Numerical results in terms of $\text{SIR}_\text{imp}$, $\text{RMSE}$ and $\text{LRMSE}$ for this scenario are presented in Table \ref{tab_exmpl1}.

Fig. \ref{res_ex1} clearly demonstrates the efficiency of the proposed methods for decontaminating the spectrum of the object of interest in this first scenario. Indeed, all methods successfully attenuate the contamination from other objects in this scenario. It should be noted that the estimated spectra are noisy, which is to be expected given the high level of noise in the observed data and the fact that our methods  aim at separating sources rather than denoising them. The results presented in Table \ref{tab_exmpl1} confirm this conclusion. It can be observed that the $\text{RMSE}$ and $\text{LRMSE}$ values for all methods are low, and the $\text{SIR}_\text{imp}$ values are high, indicating their effectiveness in removing contaminants.
 Finally, it can be seen that our {\it LC-LCMP} method yields the best results for this scenario.

\begin{table}[h]
	\centering
	\caption{ $\text{SIR}_\text{imp}$ in decibel ({\rm dB}), $\text{RMSE}$ and $\text{LRMSE}$ obtained for the first scenario.}
	\begin{tabular}{|c|c|c|c|c|c|}
		\hline	\cline{2-4} \multicolumn{1}{|c|}{Method}  & \multicolumn{1}{|c|}{ $\text{SIR}_\text{imp}$ ({\rm dB})}& \multicolumn{1}{|c|}{ $\text{RMSE}$} & \multicolumn{1}{|c|}{$\text{LRMSE}$}\\
		\hline
		{\it LI-LSQ }    & $8.88$ & $0.12$  & $0.12$ \\ 
		\hline			
		{\it LI-LSQN} & $10.15$  & $0.10$  & $0.12$ \\
		\hline
		{\it LI-MPDR }   & $9.91$ & $0.10$  & $0.12$      \\ 
		\hline		
		{\it LC-LCMP}	 & {\bf 12.35} & {\bf 0.09} & {\bf 0.11}    \\
		
		\hline
	\end{tabular}
	
	\label{tab_exmpl1}
\end{table}
\subsection{Scenario with a few hot pixels}
\begin{figure}[h!]
	\centering
	\hspace*{-0.54cm}
\begin{tikzpicture}[scale=0.5]	
\node  at (-0.5,2.2) {\includegraphics[angle=0,width=1.1\hsize]{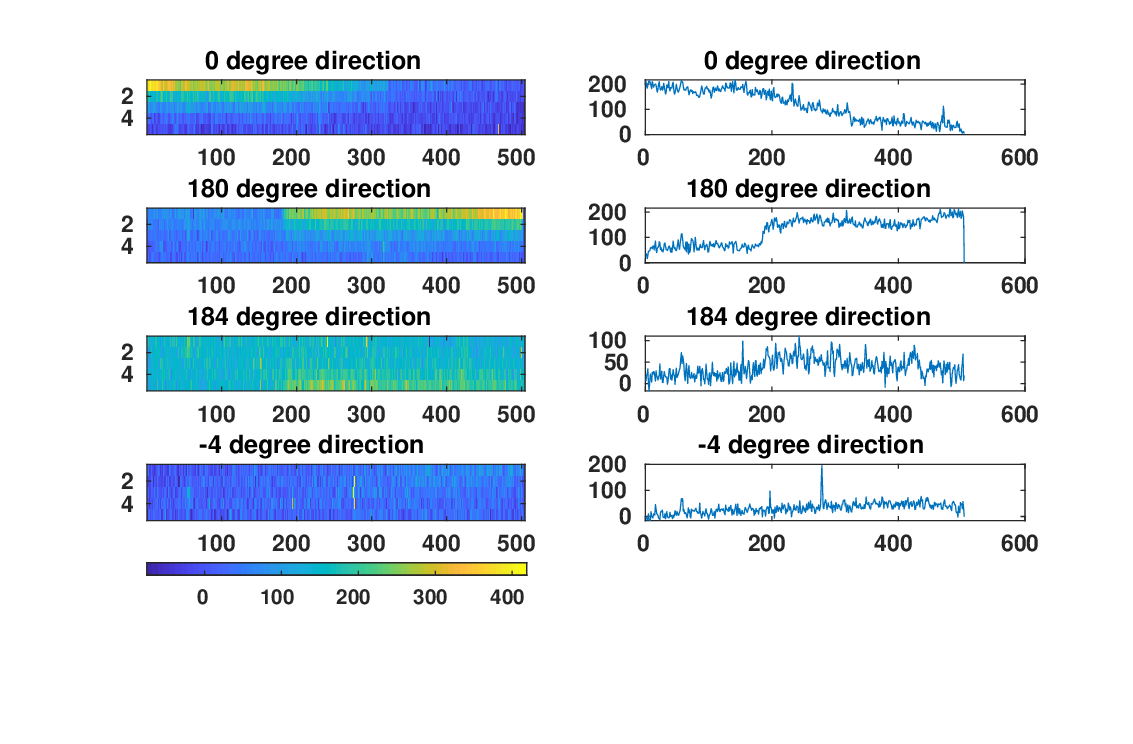}};
\node  at (4.2,-1.3) {\tiny{pixel index}};
\node [rotate=90]  at (-0.4,6.9) {\scalebox{.5}{electrons}}; 
\node [rotate=90]  at (-0.4,4.6) {\scalebox{.5}{electrons}};
\node [rotate=90]  at (-0.4,2.3) {\scalebox{.5}{electrons}}; 
\node [rotate=90]  at (-0.4,-0.0) {\scalebox{.5}{electrons}}; 
\node [rotate=90]  at (-8.6,6.9) {\scalebox{.65}{pixels}}; 
\node [rotate=90]  at (-8.6,4.6) {\scalebox{.65}{pixels}};
\node [rotate=90]  at (-8.6,2.3) {\scalebox{.65}{pixels}}; 
\node [rotate=90]  at (-8.6,-0.0) {\scalebox{.65}{pixels}}; 	 
\node  at (-4.7,-1.00) {\scalebox{.5}{pixels}};	 	 	 	 	 
\end{tikzpicture}
\vspace*{-1.5cm}
	\caption{Observed spectrograms (left) and corresponding 1D spectra (right) for the second scenario.}
	\label{observations10}
\end{figure}
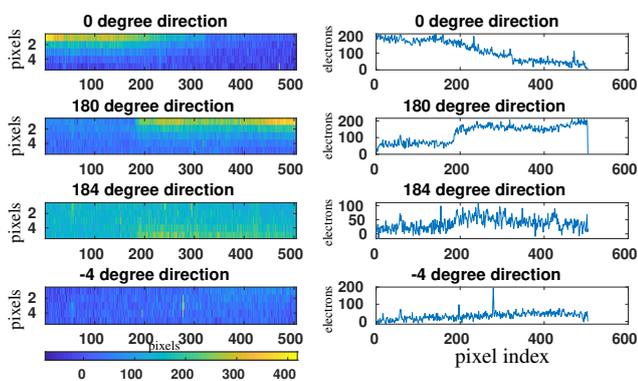

In our second experiment, we chose a complicated scenario where the spectrogram of an object of interest (a star-forming  galaxy with a redshift of $0.93$) is highly contaminated in all four directions of dispersion. In addition, there are a few hot pixels\footnote{A hot pixel is a pixel in the spectrogram that has a very high magnitude relative to neighboring pixels. These hot pixels are often the result of defects at this pixel in the detector \citep{Dokkum_2024}.} in the $-4$ degree dispersion direction.
The observed spectrograms and corresponding 1D spectra for this object are shown in Fig. \ref{observations10}. 

\begin{figure}[h!]
	\centering
	\hspace*{-0.4cm}
	\includegraphics[width=10.25cm, keepaspectratio]{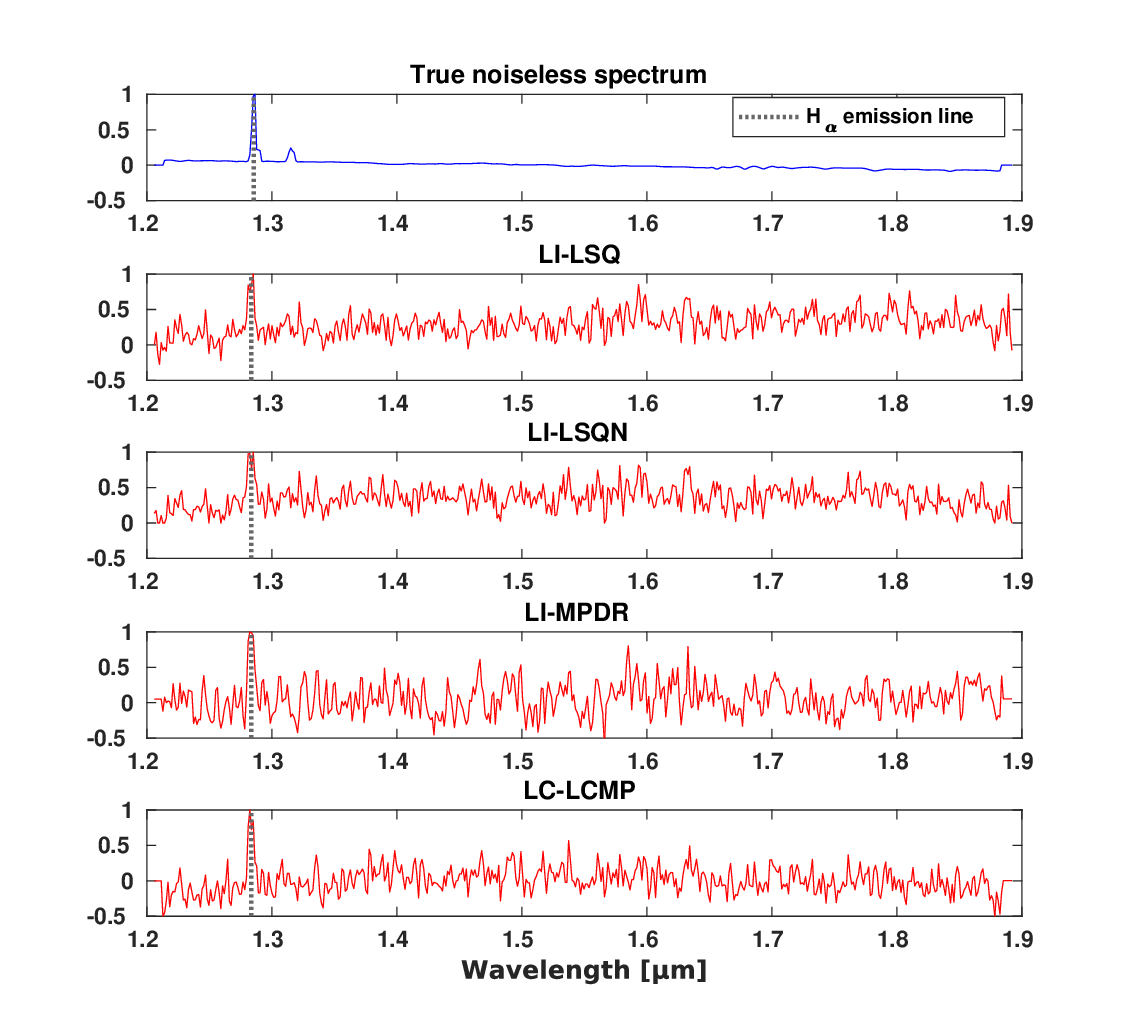}
	\vspace*{-0.5cm}
	\caption{True noiseless spectrum of the object of interest (in blue) and its estimates obtained using the proposed methods (in red) for the second scenario.}
\label{res_ex10}
\end{figure}

Fig. \ref{res_ex10} shows the true noiseless spectrum of the object of interest, along with its estimates using our methods {\it LI-LSQ}, {\it LI-LSQN}, {\it LI-MPDR} and {\it LC-LCMP} after normalization by their maximum. This figure clearly demonstrates the efficiency of the proposed methods for decontaminating the spectrum of the object of interest in this scenario.  Indeed, all of the proposed methods were successful in eliminating contamination from other objects and hot pixels, and in recovering the main emission line.

The numerical results in terms of $\text{SIR}_\text{imp}$, $\text{RMSE}$ and $\text{LRMSE}$ for this scenario are presented in Table \ref{tab_exmpl10}. This table clearly demonstrates that the {\it LC-LCMP} method is the most effective for decontaminating the spectrum of the object of interest in this scenario.
Indeed, it outperforms the {\it LI-LSQ}, {\it LI-LSQN} and {\it LI-MPDR} methods by $6.10\, {\rm dB}$, $4.64\, {\rm dB}$ and $2.45\, {\rm dB}$ respectively in terms of $\text{SIR}_\text{imp}$.
Moreover, the  $\text{RMSE}$ and the $\text{LRMSE}$ of the {\it LC-LCMP} method are lower than those of all the other methods.

\begin{table}[h!]
	\centering
	\caption{  $\text{SIR}_\text{imp}$ ({\rm dB}), $\text{RMSE}$ and $\text{LRMSE}$  obtained for the second scenario.}
	\begin{tabular}{|c|c|c|c|c|c|}
			\hline	\cline{2-4} \multicolumn{1}{|c|}{Method}  & \multicolumn{1}{|c|}{ $\text{SIR}_\text{imp}$ ({\rm dB})}& \multicolumn{1}{|c|}{ $\text{RMSE}$} & \multicolumn{1}{|c|}{$\text{LRMSE}$}\\
		\hline
		{\it LI-LSQ}    & 9.50 & $0.30$   & $0.24$ \\ 
		\hline			
		{\it LI-LSQN}& 10.36  & $0.28$  & $0.24$ \\
		\hline
		{\it LI-MPDR}   & 13.15 & $0.24$  & $0.21$     \\ 
		\hline	
		{\it LC-LCMP}	 & {\bf 15.60} & {\bf 0.19} & \boldmath{$0.17$}    \\
		
		\hline
	\end{tabular}
	
	\label{tab_exmpl10}
\end{table}

\subsection{Difficult scenario with many hot pixels}

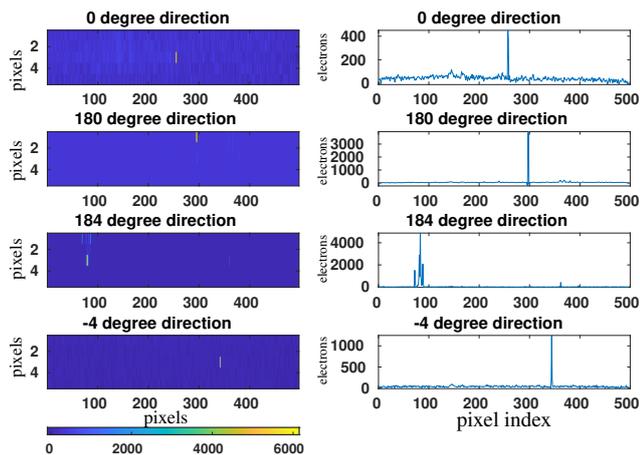
\begin{figure}[h!]
	\centering
	\hspace*{-0.44cm}
\begin{tikzpicture}[scale=0.5]	
\node  at (-0.5,2.2) {\includegraphics[angle=0,width=1.1\hsize]{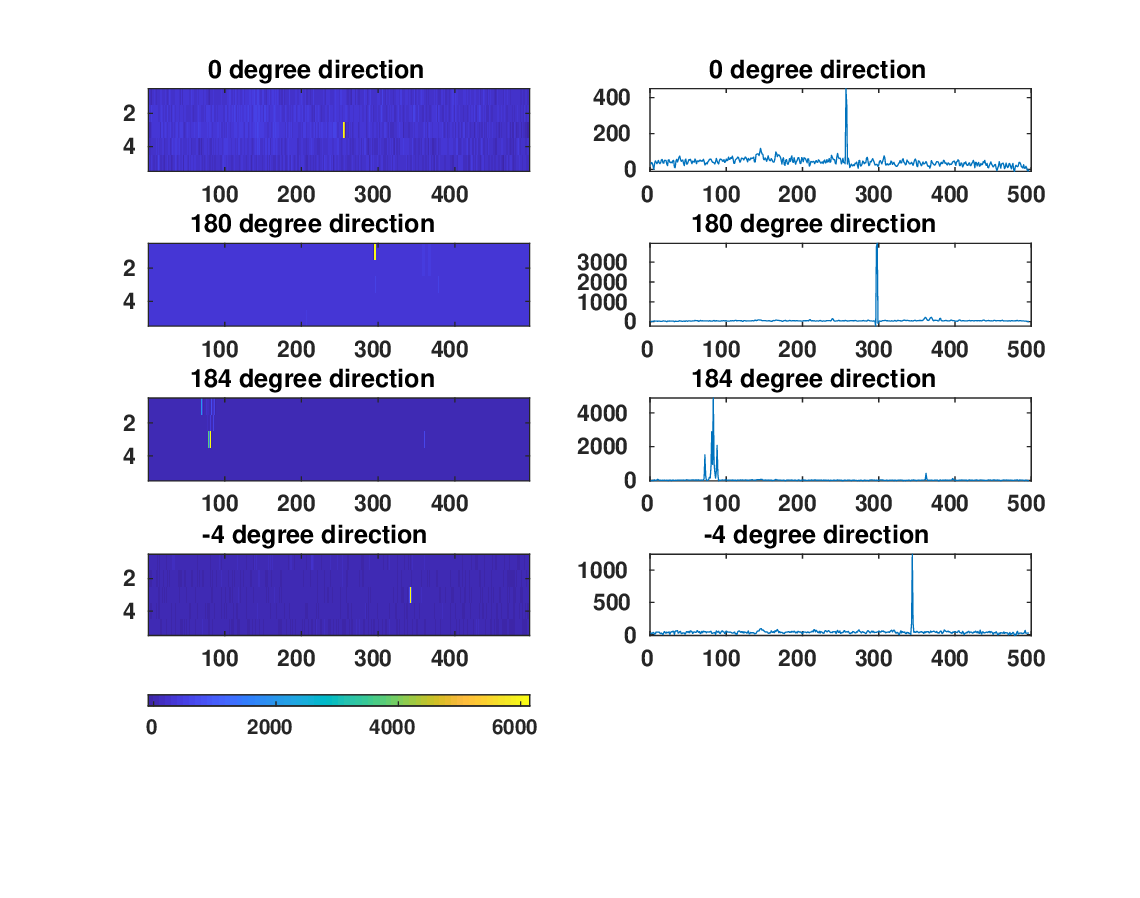}};
\node  at (4.2,-1.9) {\tiny{pixel index}};
\node [rotate=90]  at (-0.58,7.75) {\scalebox{.5}{electrons}}; 
\node [rotate=90]  at (-0.58,5.0) {\scalebox{.5}{electrons}};
\node [rotate=90]  at (-0.58,2.4) {\scalebox{.5}{electrons}}; 
\node [rotate=90]  at (-0.58,-0.3) {\scalebox{.5}{electrons}}; 
\node [rotate=90]  at (-8.6,7.4) {\scalebox{.65}{pixels}}; 
\node [rotate=90]  at (-8.6,5.0) {\scalebox{.65}{pixels}};
\node [rotate=90]  at (-8.6,2.4) {\scalebox{.65}{pixels}}; 
\node [rotate=90]  at (-8.6,-0.2) {\scalebox{.65}{pixels}}; 	 
\node  at (-4.7,-1.84) {\scalebox{.7}{pixels}};	 	 	 	 	 
\end{tikzpicture}
\vspace*{-1.5cm}
	\caption{Observed spectrograms (left) and corresponding 1D spectra (right) for the third scenario.}
	\label{observations2}
\end{figure}

In our third experiment, we chose a difficult scenario where the spectrogram of an object of interest (a star-forming galaxy with a redshift of $1.12$) is highly contaminated in all four dispersion directions. Additionally, there is at least one hot pixel present in each dispersion direction, further complicating this scenario. 
The observed spectrograms and corresponding 1D spectra for this object are presented in Fig. \ref{observations2}, where the isolated peaks in the observed 1D spectra are due to 
hot pixels. Given that the true position of the main $H_\alpha$ emission line is around the wavelength with index $145$, it cannot be distinguished in the spectra of Fig. \ref{observations2} due to the strong contribution of noise and contaminants.

\begin{figure}[h!]
	\centering
	\hspace*{-0.75cm}
	\includegraphics[width=10.25cm, keepaspectratio]{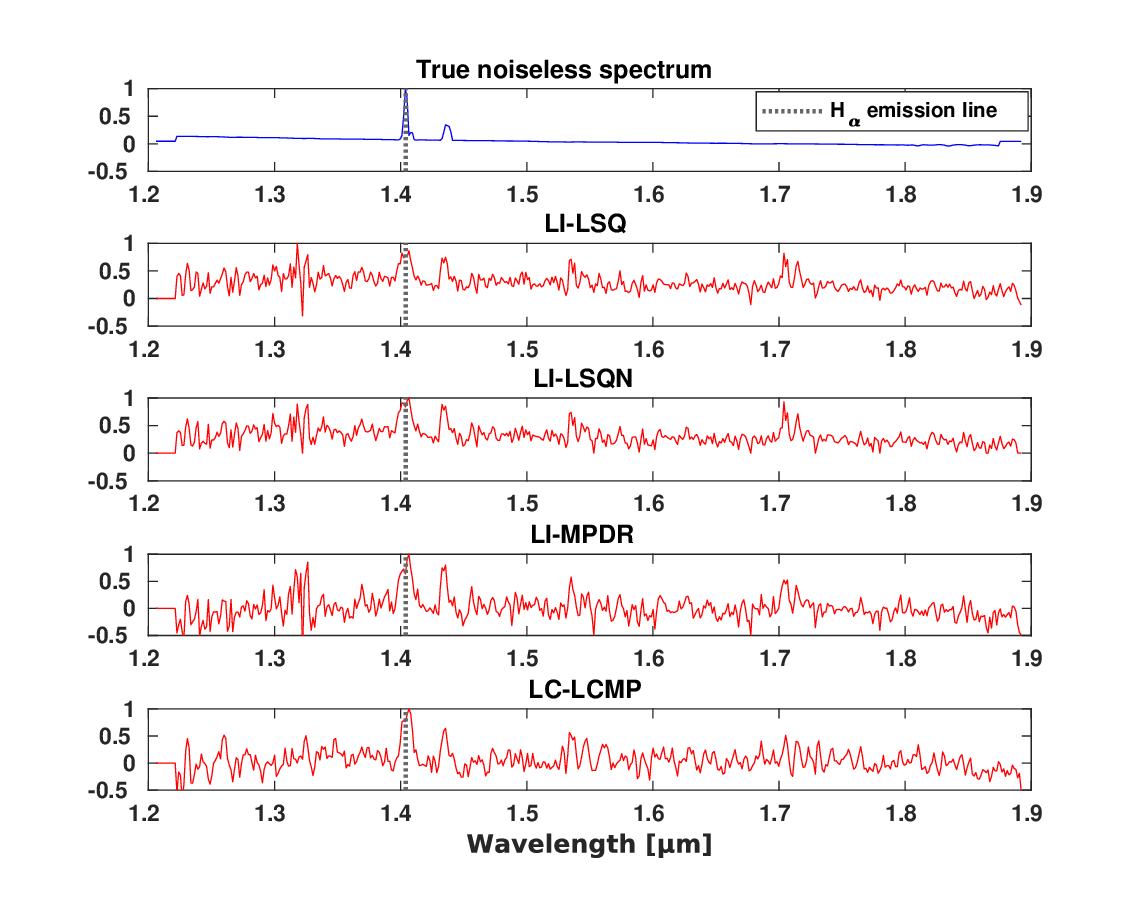}
	\vspace*{-0.5cm}
	\caption{True noiseless spectrum of the object of interest (in blue) and its estimates obtained using the proposed methods (in red) for the third scenario.}
\label{res_ex2}
\end{figure}

Fig. \ref{res_ex2} shows the true noiseless spectrum of the object of interest and its estimates using the above methods after normalization by their maximum. 
This figure clearly demonstrates the efficiency of the proposed methods {\it LI-MPDR} and {\it LC-LCMP} for decontaminating the spectrum of the object of interest in this challenging scenario.
This result is expected because these two methods use beamforming techniques, which should perform better in the  presence of unmodeled interferences such as hot pixels.

Table \ref{tab_exmpl2} clearly demonstrates that the {\it LC-LCMP} method is the most effective for decontaminating the spectrum of the object of interest in this scenario. Indeed, it outperforms the {\it LI-LSQ}, {\it LI-LSQN} and {\it LI-MPDR} methods by $6.48\, {\rm dB}$, $5.75\, {\rm dB}$ and $1.59\, {\rm dB}$ respectively in terms of $\text{SIR}_\text{imp}$. Moreover, the  $\text{RMSE}$ and the $\text{LRMSE}$ of the {\it LC-LCMP} method are lower than those of all other methods, validating the effectiveness of this method in this third scenario.

\begin{table}[h]
	\centering
	\caption{ $\text{SIR}_\text{imp}$ ({\rm dB}), $\text{RMSE}$ and $\text{LRMSE}$  obtained for the third scenario.}
	\begin{tabular}{|c|c|c|c|c|c|}
		\hline	\cline{2-4} \multicolumn{1}{|c|}{Method}  & \multicolumn{1}{|c|}{ $\text{SIR}_\text{imp}$ ({\rm dB})}& \multicolumn{1}{|c|}{ $\text{RMSE}$} & \multicolumn{1}{|c|}{$\text{LRMSE}$}\\
		\hline
		{\it LI-LSQ}    & $12.63$ & $0.36$  & $0.20$ \\ 
		\hline			
		{\it LI-LSQN} & $13.36$ & $0.31$   & $0.19$ \\
		\hline
		{\it LI-MPDR}   & $17.52$ & $0.22$ & $0.17$      \\ 
		\hline		
		{\it LC-LCMP}	 & {\bf 19.11} &{\bf 0.19} & {\bf 0.16}    \\
		
		\hline
	\end{tabular}
	
	\label{tab_exmpl2}
\end{table}
\subsection{Performance using 100 objects}

To confirm the effectiveness of the proposed methods, we performed the same  tests on a total of 100 different objects. The  chosen objects for these tests have a redshift $0.9<z<1.8$, an $H_\alpha$ flux greater than $10^{-16}$ ${\rm erg/s/cm^2}$ and a magnitude varying between $20$  and $23.5$  AB mag.
For the comparison, we used the NMF Alternating Least Square (ALS) method ({\it NMF-ALS}) described in \citep{Paatero}. 

The {\it NMF-ALS} method is a BSS method based on source positivity. It involves finding an approximate decomposition of the total observation matrix ${\bf X}$, as defined in (\ref{Mix_LI}), into two non-negative matrices, with the aim of identifying one of these matrices as the mixing matrix ${\bf A}$ and the other as the source matrix ${\bf E}$.
This method uses only the observation matrix ${\bf X}$ and does not utilize any additional available information, such as direct images.
 The primary limitation with the {\it NMF-ALS} method is the non-uniqueness of its solutions \citep{Paatero}. Indeed, this iterative approach may converge to the desired solution $({\bf A}, {\bf E})$, but it can also converge to a different (local) minimum that deviates from this solution.

The results, in terms of the averages of $\text{SIR}_\text{imp}$ ({\rm dB}), $\text{RMSE}$ and $\text{LRMSE}$ as well as their respective standard deviations, are presented in Table \ref{tab_comp}.

\begin{table}[tp!]
	\centering
	\caption{ Means and standard deviations of $\text{SIR}_\text{imp}$ ({\rm dB}), $\text{RMSE}$ and $\text{LRMSE}$ using 100 objects.}
	\begin{tabular}{|c|c|c|c|c|c|c|c|}
		\cline{2-7} \multicolumn{1}{c}{} & \multicolumn{2}{|c|}{ $\text{SIR}_\text{imp}$ ({\rm dB})} & \multicolumn{2}{|c|}{$\text{RMSE}$}& \multicolumn{2}{c|}{$\text{LRMSE}$ } \\
		\hline	\cline{2-7} \multicolumn{1}{|c|}{Method}  & mean  & std  & mean  & std   & mean     &   std   \\
		\hline
		{\it LI-LSQ }    & 8.08 & 5.73 & 0.36& 0.17 & $0.27$& $0.17$\\ 
		\hline			
		{\it LI-LSQN} & 8.38 & 5.82 & 0.32& 0.16 & $0.25$ & $0.16$\\
		\hline
		{\it LI-MPDR  }   & 9.36 & 6.04 & 0.30& 0.18 & $0.22$   & $0.15$ \\ 
		\hline

		{\it LC-LCMP}	 & {\bf 10.75} &   6.51 & {\bf 0.26}& 0.17 & {\bf 0.20}  &  $0.14$ \\		
		\hline	
		{\it NMF-ALS}   & 2.19 & 3.94& 0.76& 0.38 & $0.57$  &  $0.28$\\
		\hline	
	\end{tabular}
	
	\label{tab_comp}
\end{table}

Table \ref{tab_comp} clearly demonstrates that the {\it LC-LCMP} method outperforms all other methods in terms of average $\text{SIR}_\text{imp}$ ({\rm dB}), $\text{RMSE}$ and  $\text{LRMSE}$. Indeed, this method shows average $\text{SIR}_\text{imp}$ improvements of $2.67\, {\rm dB}$, $2.37\, {\rm dB}$, 1.39$\, {\rm dB}$ and 8.59$\, {\rm dB}$ compared to the {\it LI-LSQ}, {\it LI-LSQN}, {\it LI-MPDR} and {\it NMF-ALS} methods, respectively.
The superiority of the {\it LC-LCMP} method in terms of $\text{SIR}_\text{imp}$ can be explained by the fact that the convolutive mixing model used by this method is more accurate than the approximate linear instantaneous model presented in Sect. \ref{sec_MLI} and used by the other methods.
The mean $\text{LRMSE}$ of the {\it LC-LCMP} method is again lower than that of all other methods.
This result is expected and can be explained by the fact that the {\it LC-LCMP} method aims to estimate deconvolved versions of the spectra, providing more accurate estimates around the main emission line. 
In contrast, all other methods provide sources convolved by a function that depends on the object profile in the dispersion direction,  which consequently impairs the  accuracy of the estimate in the region around the main emission line.   Moreover, the mean  $\text{RMSE}$ of the {\it LC-LCMP} method is lower than that of all other methods, validating the effectiveness of this method.

In terms of standard deviations, the {\it NMF-ALS} method has a significantly lower $\text{SIR}_\text{imp}$ standard deviation than the others. The reason for this result is that this method typically produces consistently poor results, which do not vary significantly from one experiment to another when compared to the proposed methods.
 It is important to highlight that the {\it LI-LSQ} method exhibits the lowest standard deviation in terms of $\text{SIR}_\text{imp}$ among all the proposed methods, while the {\it LI-LSQN} method has the lowest standard deviation in terms of $\text{RMSE}$, and the {\it LC-LCMP} method shows the lowest standard deviation in terms of $\text{LRMSE}$.

\begin{figure*}[h!]
	\centering
	\subfloat{\includegraphics[width=0.33\textwidth,keepaspectratio]{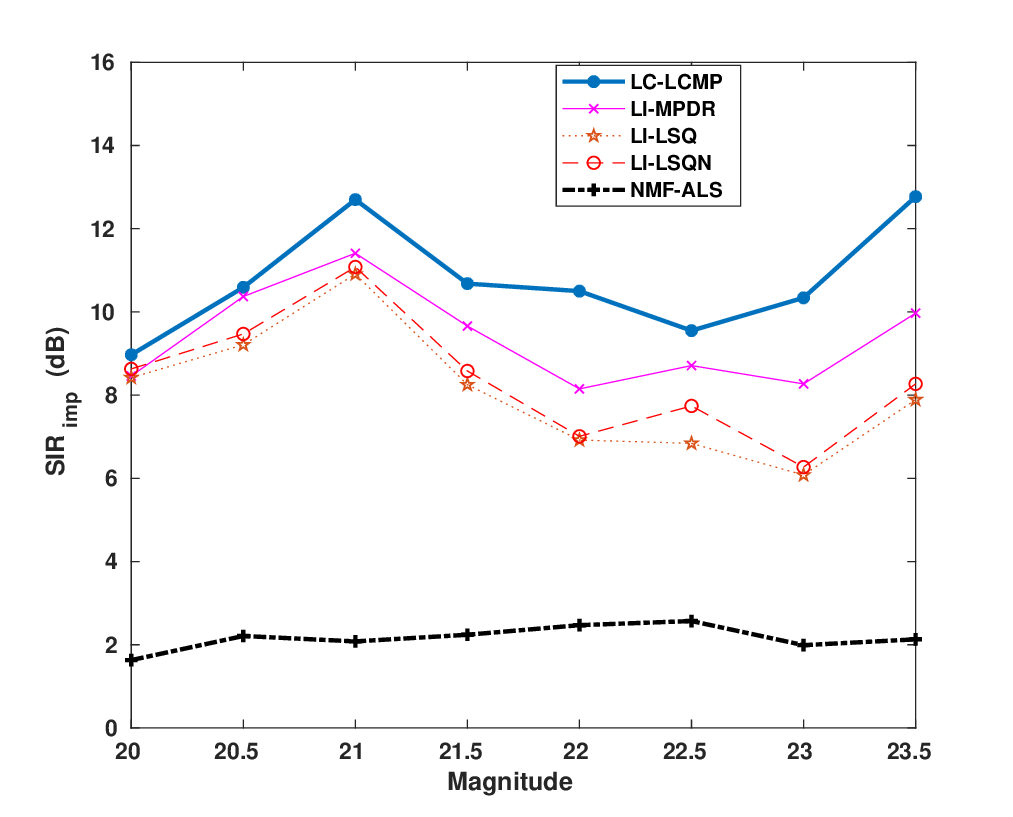}}
	\subfloat{\includegraphics[width=0.33\textwidth,keepaspectratio]{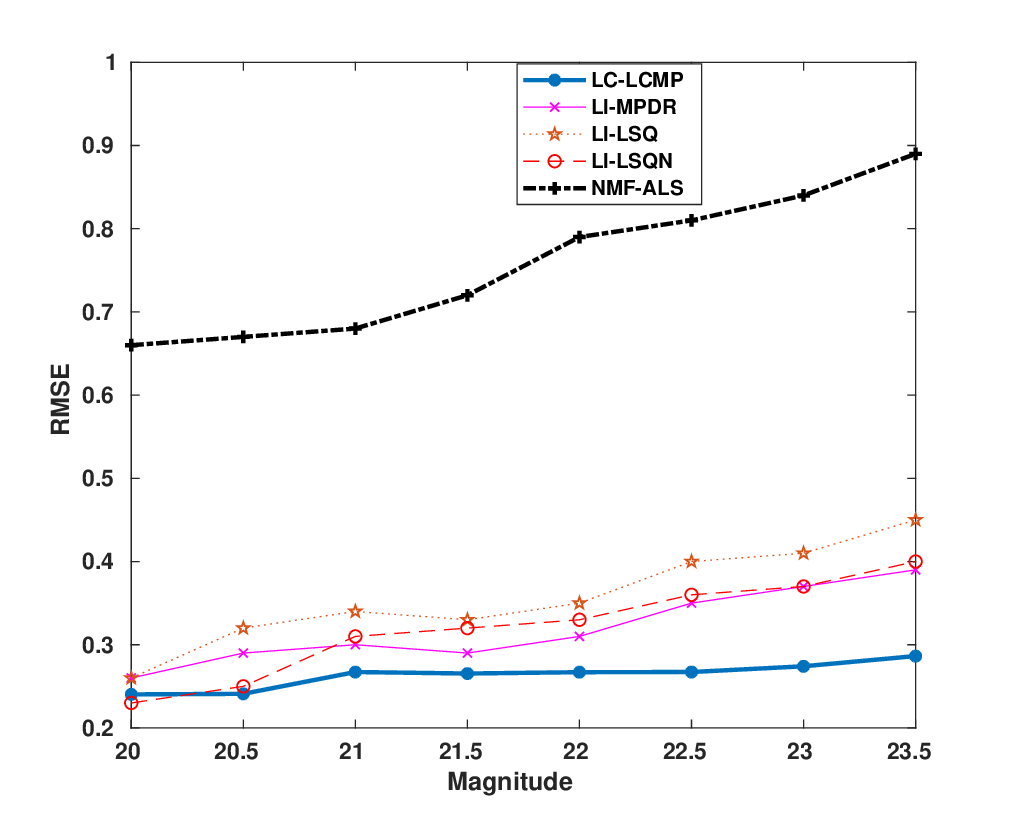}}
	\subfloat{\includegraphics[width=0.33\textwidth,keepaspectratio]{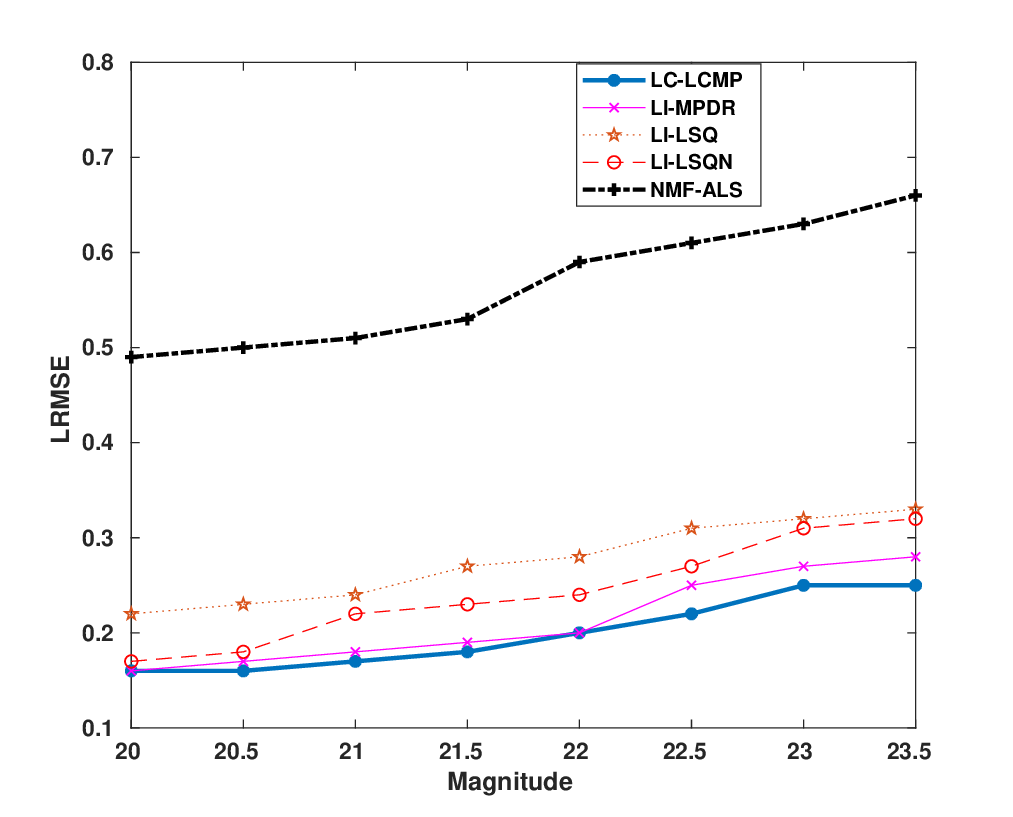}}
	
	\caption{Performance of the methods as a function of magnitude.}
	\label{res_mag}
\end{figure*}

\begin{figure*}[h!]
	\centering
	\subfloat{\includegraphics[width=0.33\textwidth,keepaspectratio]{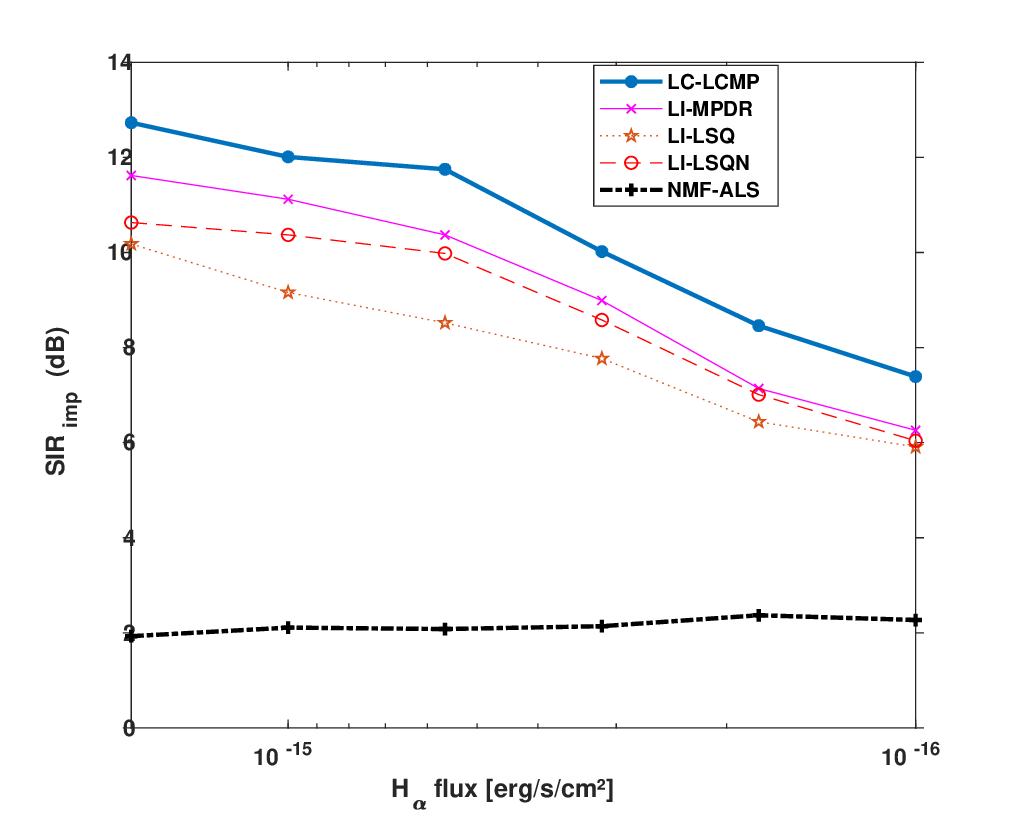}}
	\subfloat{\includegraphics[width=0.33\textwidth, keepaspectratio]{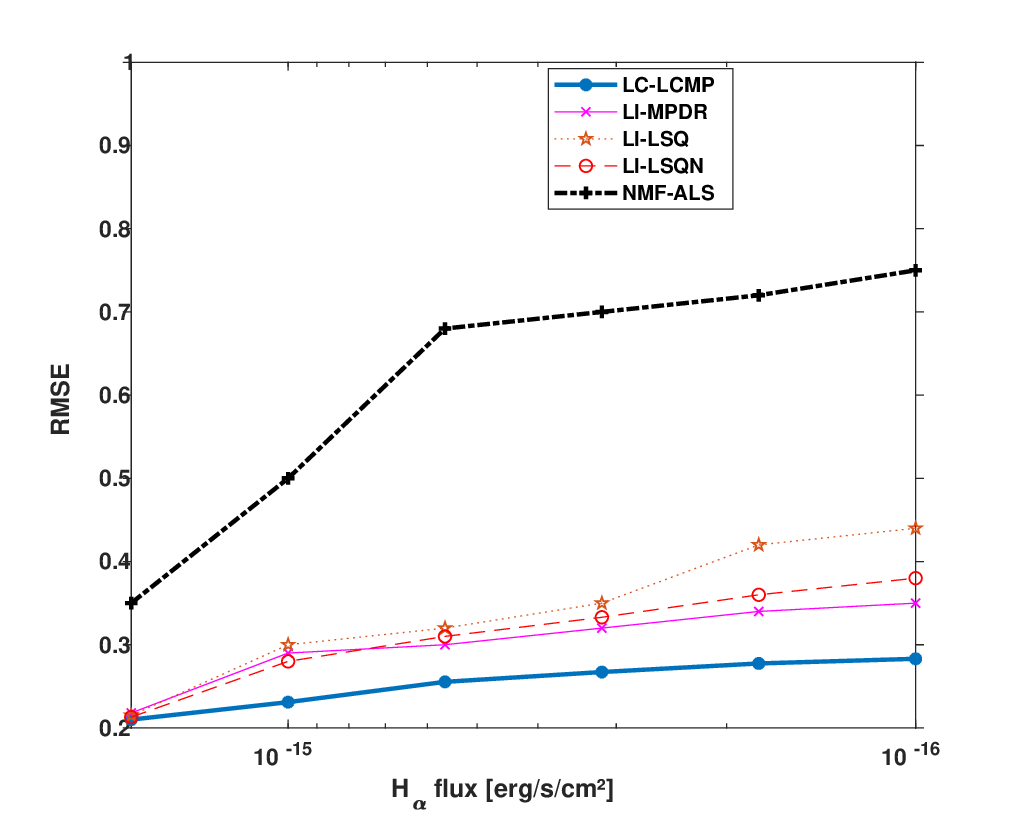}}
	\subfloat{\includegraphics[width=0.33\textwidth, keepaspectratio]{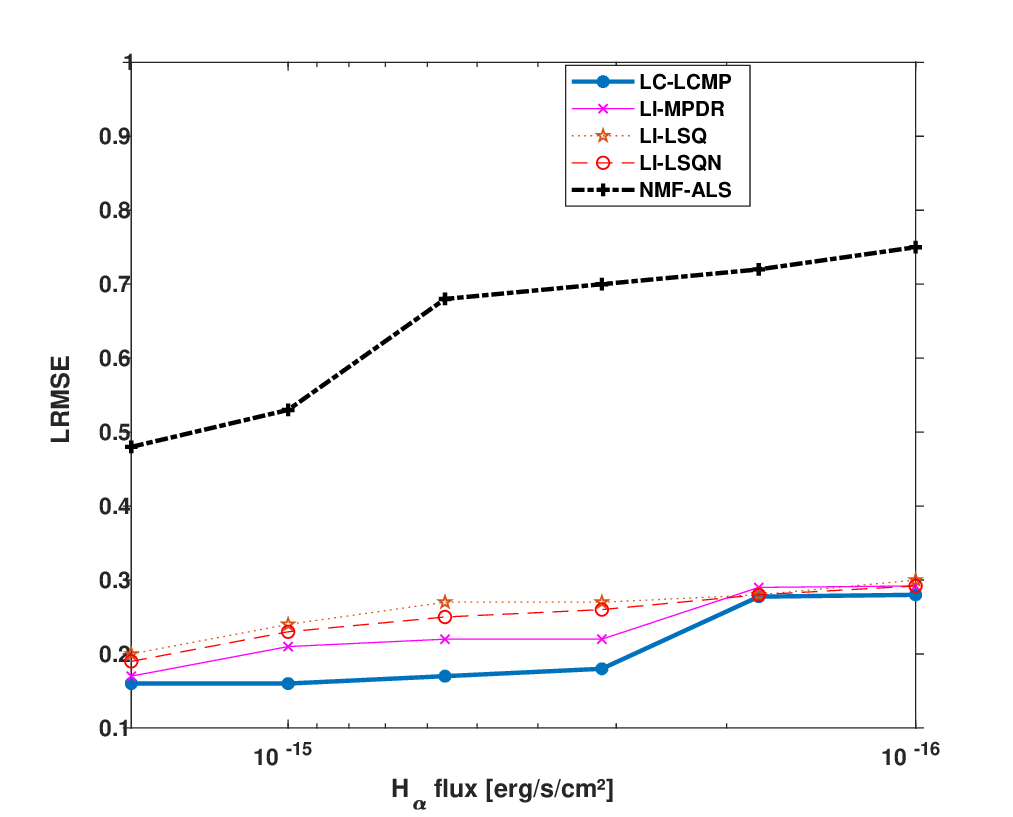}}
	
	\caption{Performance of the methods as a function of $H_\alpha$ flux.}
	\label{res_flux}
\end{figure*}

Next, we examine the behavior of our methods as a function of object magnitude and  $H_\alpha$ flux, using these 100 objects. Figs. \ref{res_mag} and \ref{res_flux} show the performance obtained in this experiment in terms of $\text{SIR}_\text{imp}$, $\text{RMSE}$ and $\text{LRMSE}$.

Fig. \ref{res_mag} shows that the performance of our methods in terms of $\text{SIR}_\text{imp}$ remains relatively stable as a function of the magnitude of the objects to be decontaminated.
Regarding the performance of the methods in terms of $\text{RMSE}$ and $\text{LRMSE}$, we generally observe that for all the methods,  an increase in the magnitude of the objects to be decontaminated leads to higher $\text{RMSE}$ and $\text{LRMSE}$ values.   This is to be expected, as the lower the brightness of an object, the more challenging it is to decontaminate it due to the increased number of contaminants that are brighter than the object.
Fig. \ref{res_flux} shows that  a decrease in the $H_\alpha$ flux of the objects to be decontaminated results in a corresponding decrease in the performance of our methods in terms of $\text{SIR}_\text{imp}$, $\text{RMSE}$ and $\text{LRMSE}$.

For all the proposed methods except for {\it LI-MPDR}, the computation time varies depending on the number of contaminants and treated objects. The {\it LI-MPDR} method offers the best performance in terms of computation time. In fact, its computation time is almost half that of the {\it LI-LSQ}, {\it LI-LSQN}, and {\it LC-LCMP} methods.

\section{Conclusion and perspectives}
\label{Conclusion}

In this paper, we proposed  four new separation methods adapted to \Euclid-like spectra and based on two different local approaches. The first approach is called the local instantaneous approach, which uses an approximate instantaneous model. The second approach is called the local convolutive approach, which uses a more realistic convolutive model.
For each approach, we have developed a mixing model that links the observed data to the source spectra, simultaneously taking into account four grism dispersion directions, either in the spatial domain for the instantaneous approach, or in the Fourier domain for the convolutive approach. We then developed several methods for decontaminating these spectra from mixtures, exploiting the direct images provided by the near-infrared photometers.
Test results obtained in Sect. \ref{test} confirm the effectiveness of all the proposed methods. In particular, the local convolutive method performs remarkably well in terms of $\text{SIR}_\text{imp}$, $\text{RMSE}$ and $\text{LRMSE}$,   clearly outperforming other methods based on a linear instantaneous mixing model.
\\

In terms of perspectives, it would be interesting to apply the proposed methods to real \Euclid satellite data when available. Additionally, replacing the local approach used by all the methods proposed in this article, where decontamination is carried out object by object, with a global approach where all objects located in a predefined zone are decontaminated simultaneously could be beneficial.
Also, in our work, we have exploited direct images of the spectral band $J$ to obtain an estimate of the mixing coefficients. When direct images of the other bands $Y$ and $H$ become available, they can also be exploited to improve our estimate of the mixing coefficients.
Finally, in the proposed models, we have assumed that the PSF is spectrally invariant, whereas this PSF varies slightly as a function of wavelength. Taking this variation into account could improve the accuracy of the proposed models.

\begin{acknowledgements}
The authors thank their colleagues from the SIR and SIM organizational units of the \Euclid project. The SIM team provided the essential simulated data used in this work, and the SIR team performed the crucial data preprocessing steps. Their contributions were foundational to this work.

\end{acknowledgements}

\bibliography{aa57144-25}

\end{document}